\documentclass[traditabstract]{aa} % for the abstract without structuration 
\usepackage{graphicx}
\usepackage{natbib}
\usepackage{txfonts}

\begin{document}

   \title{An investigation of star formation and dust attenuation in major mergers using ultraviolet and infrared data}

   \author{F.-T. Yuan
          \inst{1}
          \and
          T. T. Takeuchi\inst{1}
		  \and
		  Y. Matsuoka\inst{1}      
          \and
          V. Buat\inst{2}
          \and           
          D. Burgarella\inst{2}
          \and
          J. Iglesias-P\'{a}ramo\inst{3, 4}        
          }

   \institute{Division of Particle and Astrophysical Sciences, Nagoya University, Furo-cho, Chikusa-ku, 464-8602, Japan\\
              \email{yuan.fangting@g.mbox.nagoya-u.ac.jp}
         \and
             Aix-Marseille  Universit\'e,  CNRS, LAM (Laboratoire d'Astrophysique de Marseille) UMR7326,  13388, France
          \and
          Instituto de Astrof\'{\i}sica de Andaluc\'{\i}a (IAA - CSIC), Glorieta de la Astronom\'{\i}a s.n., 18008 Granada, Spain
          \and
          Centro Astron\'{o}mico Hispano Alem\'{a}n, C/ Jes\'{u}s Durb\'{a}n Rem\'{o}n 2-2, 04004 Almer\'{\i}a, SPAIN
             }

   \date{}

% \abstract{}{}{}{}{} 
% 5 {} token are mandatory
 
  \abstract{Merger processes play an important role in galaxy formation and evolution. To study the influence of merger processes on the evolution of dust properties and cosmic star formation rate, we investigate a local sample of major merger galaxies and a control sample of isolated galaxies using GALEX ultraviolet (UV) and Spitzer infrared (IR) images. Through a statistical study, we find that dust attenuation in merger galaxies is enhanced with respect to isolated galaxies. We find this enhancement is contributed mainly by spiral galaxies in spiral-spiral (S-S) pairs, and increases with the increasing stellar mass of a galaxy. Combining the IR and UV parts of star formation rates (SFRs), we then calculated the total SFRs and specific star formation rates (SSFRs). We find the SSFRs to be enhanced in merger galaxies. This enhancement depends on galaxy stellar mass and the companion's morphology, but depends little on whether the galaxy is a primary or secondary component or on the separation between two components. These results are consistent with a previous study based only on IR images. In addition, we investigate the nuclear contributions to SFRs. SFRs in paired galaxies are more concentrated in the central part of the galaxies than in isolate galaxies. Our studies of dust attenuation show that the nuclear parts of pairs most resemble ULIRGs. Including UV data in the present work not only provides reliable information on dust attenuation, but also refines analyses of SFRs. }


   \authorrunning{Yuan et al.}
   \titlerunning{SFRs and dust attenuation of mergers}
   \maketitle
%
%________________________________________________________________

\section{Introduction}
\label{sec_intro}

In the hierarchical scenario of galaxy and structure formation, interactions between galaxies and their associated dark matter halos happen frequently \citep[e.g][]{cole_hierarchical_2000, wechsler_concentrations_2002, li_assembly_2007, freedman_woods_triggered_2010} and can strongly affect galaxy properties, such as morphology, luminosity, star formation rate (SFR), and dust properties \citep[e.g.][]{struck_galaxy_2006, hwang_goods-herschel:_2011}. It is therefore quite important to consider the effects of galaxy-galaxy interaction when studying the evolution of galaxies. 

If the orbital energy of the two interacting galaxies is low enough, merging of galaxies occurs, and these mergers play a very important role in the formation and evolution of galaxies and their dark matter halos. Simulations show that during the merging process of gas-rich galaxies, the gas flows inward and causes a starburst in the nuclear region \citep[e.g.][]{toomre_galactic_1972}. Although mergers are thought to be connected with starburst and AGN activity, the details are still not clear, and several key questions related to mergers still remain to be solved \citep{mo_galaxy_2010}. 

One of the questions is related to the strong evolution of the cosmic star formation density from $z=0$ to $z=1$ \citep[e.g.][]{lilly_canada-france_1996, madau_star_1998, hopkins_evolution_2004, takeuchi_evolution_2005}. This evolution can be caused by a change of merger rate at higher redshift \citep{zheng_hst/wfpc2_2004, hammer_did_2005, bridge_role_2007}. However, this conclusion is objected to in several works, which have found that the merger rate does not evolve much from $z\sim 0$ to $z \sim 1$ and that the properties that do not strongly affect galaxy morphology should be responsible for the evolution \citep{flores_15_1999, bell_toward_2005, melbourne_optical_2005, lotz_evolution_2008}. %Therefore, a quantitative study is still necessary to solve this problem.

The discrepancy between these works may be caused by the different methods they applied to selecting mergers \citep{xu_local_2010}. There are two common methods to select merger galaxies, both with advantages and disadvantages. One is to select peculiar galaxies. This method can select galaxies at a late stage of merging to very high redshift with the high-resolution images provided by large telescopes (e.g {\it HST}). However, it is uncertain whether all peculiar galaxies are in a merging state, since some of them are isolated galaxies showing irregular star formation regions. The patchy distribution of dust in these regions can strongly affect the light distribution and make this method of identification problematic, especially in ultraviolet (UV) bands \citep{burgarella_2001}. Another disadvantage of this method is that the signature of the disturbed features becomes harder to detect as one goes deeper in space \citep{mo_galaxy_2010, xu_local_2010}. 

The other method is to select close pairs, assuming that these pairs will ultimately merge within a certain time scale. This method avoids the complex identification of morphological features, and thus is more objective. However, it suffers from biases such as (1) the projection effect, (2) lack of very close pairs owing to the limited resolution of the telescope, (3) lack of less luminous components given a certain magnitude limit (`missing the secondary') \citep{xu_near-infrared_2004, xu_local_2010}.     

To reduce these biases, \cite{xu_local_2010} (hereafter Xu10) built a local sample of close pairs selected from near-infrared bands by carefully choosing the selection criteria and presented Spitzer observations for these galaxies. They find an apparent enhancement of star formation rates in mergers and discuss in detail the dependence of the enhancement on morphology, mass, and separation of these galaxies. Their work is less affected by dust extinction than are previous studies based on UV and optical bands \citep[e.g.][]{barton_tidally_2000, woods_tidally_2006, ellison_galaxy_2008}, and it has better resolution than previous infrared (IR) works \citep[e.g.][]{kennicutt_effects_1987, telesco_enhancement_1988, xu_solentic_1991}. Compared with \cite{smith_spitzer_2007}, who looked into interacting features such as bridges and tails in resolved pairs using Spitzer data, Xu10 sampled galaxies with and without strong interacting features. 

Xu10's sample provides a good opportunity to study the physical properties of local merger galaxies. One important property is dust attenuation. It changes the spectral energy distribution of a galaxy and is correlated with SFR \citep[e.g][]{burgarella_ultraviolet--far_2006, iglesias-paramo_star_2006, garn_best_2010}. Dust attenuation in major mergers can be quite complex and different from that in isolated galaxies: merging processes can inject gas into a galaxy and enhance the gas content \citep{hernquist_barnes_1991, barnes_hernquist_1996}, but it can also dissipate gas through hydrodynamic effects \citep{park_choi_2009}. Previous works about dust attenuation in merger galaxies are mostly based on theoretical modeling and numerical simulations \citep[e.g.][]{bekki_degree_2000, goldader_far-infrared_2002}. \cite{hwang_goods-herschel:_2011} study dust properties using Herschel data, but their work is confined to IR bands. Since dust attenuation of galaxies can be accurately investigated using the fraction of IR to UV luminosity $L_{\rm IR}/L_{\rm UV}$ \citep[e.g.][]{meurer_dust_1999, buat_dust_2005, burgarella_ultraviolet--far_2006, cortese_ultraviolet_2008, boquien_irx-_2012}, we aim to combine UV and IR data to statistically investigate the dust attenuation of major merger galaxies. 

The dust attenuation can be calculated without using IR data if the IRX-$\beta$ relation (relation between $L_{\rm IR}/L_{\rm UV}$ and the slope of UV spectra $\beta$) applies. This relation was originally found by \cite{meurer_dust_1999} for starburst galaxies. Later studies show that this relation changes for more quiescent galaxies \citep[e.g.][etc.]{kong_star_2004, cortese_ultraviolet_2008, munoz-mateos_dust_2009}. Recently, \cite{takeuchi2012} have found that the original IRX-$\beta$ relation needs to be corrected for the aperture effect. We examine the application of this relation to major merger galaxies in this work.

If dust attenuation in merger galaxies differs from isolated galaxies statistically, Xu10's result for SFRs may be biased because they only used IR data and ignored the effect of unobscured SFRs that can be obtained from UV observations. \cite{takeuchi_evolution_2005} have shown that the SFRs indicated by UV emission contribute about $50\%$ of the cosmic star formation density in the local universe, implying that unobscured SFRs are quite important for local galaxies. To give an unbiased view of the SFR, it is necessary to consider both the obscured and unobscured emission from young stars, namely, IR and UV emission \citep[see][etc.]{buat_star_1996, hirashita_star_2003, takeuchi_star_2010}. \cite{hancock_probing_2008} studied the UV and mid-IR properties of interacting galaxies, but they only focused on the single case of Arp 82 and its merging features. 

In this paper, we investigate the dust attenuation of paired galaxies from Xu10's sample, and re-estimate the SFRs of paired galaxies by combining the IR and UV indicated SFRs. The structure of our paper is as follows. First, we introduce the sample selection and the flux estimation of IR and UV images in Section \ref{sec_data}. Then we discuss the results of dust attenuation, the dependence of the attenuation on different parameters, and the IRX-$\beta$ relation in Section \ref{sec_dustatt}. In Section \ref{sec_ssfr} we calculate the SFRs and specific star formation rates (SSFRs) indicated by UV and IR, and compare the results with previous works. The nuclear contributions to SFRs are also discussed. Section \ref{sec_importance} is a discussion of the importance of including the unobscured SFRs. Our summary and conclusion are given in Section \ref{sec_conclusion}. Throughout this paper, the $\Lambda$-cosmology is adopted: $\Omega_{m}=0.3$, $\Omega_{\Lambda}=0.7$, and $H_{0}=75\ {\rm km\,s^{-1}\,Mpc^{-1}}$. 

%__________________________________________________________________

\section{Data}
\label{sec_data}

\subsection{Pair selection}
\label{subsec_sel}

%                                     Two column figure (place early!)
%______________________________________________ 

\begin{figure*}
\centering
\includegraphics[width=12cm]{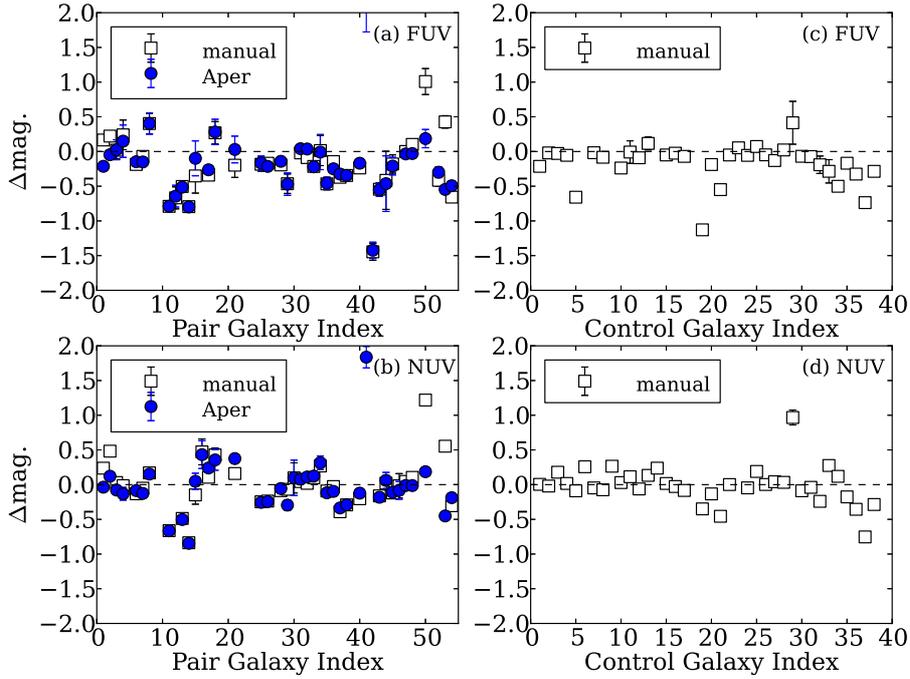}
\caption{Comparison between different methods of photometry. Panels (a) and (b) are for paired galaxies. Panels (c) and (d) are for control galaxies. Squares indicate the magnitude difference between manual measurements and GALEX pipeline data, and dots show the difference between aperture measurements and GALEX pipeline data. \label{fig_comp_pho}}
\end{figure*}

The pair and control samples were adopted from Xu10 (Table \ref{tab_add}). The galaxy pairs were selected from cross-matches between the 2MASS {\it K$_{s}$} Extended Source Catalog (XSC; \citealt{jarrett_2mass_2000}) and the galaxy catalog of SDSS-DR3 \citep{abazajian_third_2005}. This sample includes all spectroscopically confirmed spiral-spiral (S-S) and spiral-elliptical (S-E) pairs in a parent sample that is complete for primaries brighter than \textsl{K}$=12.5$ mag, projected separations between $5 h^{-1}$ kpc and $20 h^{-1}$ kpc, and mass ratio $\leq$ 2.5. There are 54 galaxies (27 pairs) and 39 of which are non-AGN spirals. The details of the selection can be found in \cite{xu_near-infrared_2004}, \cite{domingue_2mass/sdss_2009}, and Xu10. 

For each non-AGN spiral galaxy in the pair sample, one isolated galaxy with a similar mass ($\Delta \log M \leq 0.1$) and the closest redshift is matched to it. The isolated galaxies must be non-AGN late type galaxies in the local Universe ($z\leq 0.1$) with both IRAC and MIPS data, and have {\it K}$_{s}$ band magnitudes less than 13.5 mag. Although there is a difference in redshift between one galaxy in the pair sample and its isolated counterpart, Xu10 have proved that the difference will not introduce any bias using Monte-Carlo simulation. 

Previous studies of S-E pairs \citep[e.g][]{de_mello_1996,domingue_2003} have found that interactions with late-type galaxies can provoke star formation activity in an early-type galaxy. Indeed, there is one elliptical galaxy (J10514368+5101195) in our sample showing significant signs of star formation. However, since the number is small, we focus on spiral galaxies in this study.

%table_add
\begin{table*}\scriptsize
\caption{Physical properties of spiral galaxies in the pair sample and their counterparts in the control sample (Xu10).}
\label{tab_add}
\begin{center}
\begin{tabular}{ccccccccc}
\hline\hline
Paired Galaxy ID & z & $\log M$ & CAT\tablefootmark{1} & SEP\tablefootmark{2} & separation & Control galaxy ID & z & $\log M$\\ 
   &   & ($M_{\odot}$) &  &  & (kpc) &  &  & ($M_{\odot}$)\\
\hline

J00202580+0049350 & 0.0176 & 10.84 & SE2 & 0.50 &9.44 & LCK-287434 & 0.0320 & 10.88  \\
J01093517+0020132 & 0.0447 & 11.05 & SE2 & 1.12 &19.02 & LCK-178064 & 0.0450 & 11.15 \\
J01183556-0013594 & 0.0475 & 10.93 & SS2 & 1.06 &20.18 & LCK-320371 & 0.0471 & 10.96  \\
J02110832-0039171 & 0.0199 & 10.98 & SS1 & 0.56 & 8.89 & LCK-523686 & 0.0452 & 10.94  \\
J09374413+0245394 & 0.0230 & 11.46 & SE1 & 0.68 &17.29 & LCK-415950 & 0.0317 & 11.37  \\
J10205188+4831096 & 0.0531 & 10.88 & SE2 & 0.88 &17.64 & LCK-086596 & 0.0470 & 10.86  \\
J10272950+0114490 & 0.0223 & 10.73 & SE2 & 0.65 &9.28  & EN1-158103 & 0.0298 & 10.70  \\
J10435053+0645466 & 0.0273 & 10.83 & SS1 & 1.27 &15.98 & EN1-360222 & 0.0429 & 10.74  \\
J10435268+0645256 & 0.0273 & 10.73 & SS2 & 1.27 &15.98 & EN1-010947 & 0.0367 & 10.67  \\
J10514450+5101303 & 0.0244 & 11.13 & SE2 & 0.15 &4.74  & LCK-162208 & 0.0240 & 11.12  \\
J12020424+5342317 & 0.0642 & 11.16 & SE2 & 0.87 &17.90 & EN1-018834 & 0.0631 & 11.06  \\
J13082964+0422045 & 0.0241 & 10.53 & SS1 & 1.29 &12.09 & LCK-233199 & 0.0269 & 10.60  \\
J13325525-0301347 & 0.0472 & 10.90 & SS2 & 0.79 &14.29 & LCK-019297 & 0.0469 & 10.96  \\
J13325655-0301395 & 0.0472 & 11.21 & SS1 & 0.79 &14.29 & LCK-703238 & 0.0444 & 11.20  \\
J13462001-0325407 & 0.0236 & 11.01 & SE1 & 1.28 &16.79 & LCK-050667 & 0.0457 & 10.92  \\
J14005782+4251207 & 0.0327 & 11.01 & SS1 & 1.37 &19.27 & LCK-027930 & 0.0458 & 11.06  \\ 
J14005882+4250427 & 0.0327 & 10.90 & SS2 & 1.37 &19.27 & LCK-071868 & 0.0466 & 10.94  \\
J14250739+0313560 & 0.0359 & 10.66 & SE2 & 1.31 &15.38 & EN1-516050 & 0.0381 & 10.66  \\
J14334683+4004512 & 0.0258 & 11.25 & SS1 & 1.22 &19.29 & LCK-641925 & 0.0272 & 11.19  \\
J14334840+4005392 & 0.0258 & 11.10 & SS2 & 1.22 &19.29 & LCK-400414 & 0.0281 & 11.12  \\
J15064391+0346364 & 0.0345 & 11.22 & SS1 & 1.10 &16.29 & LCK-534543 & 0.0314 & 11.14  \\
J15064579+0346214 & 0.0345 & 11.17 & SS2 & 1.10 &16.29 & LCK-136060 & 0.0483 & 11.08  \\
J15101587+5810425 & 0.0312 & 11.02 & SS1 & 0.53 &7.85  & LCK-172179 & 0.0461 & 11.02  \\
J15101776+5810375 & 0.0312 & 10.79 & SS2 & 0.53 &7.85  & LCK-564807 & 0.0446 & 10.76  \\
J15281276+4255474 & 0.0182 & 11.26 & SS1 & 1.32 &17.68 & LCK-621286 & 0.0454 & 11.32  \\
J15281667+4256384 & 0.0182 & 11.03 & SS2 & 1.32 &17.68 & LCK-038716 & 0.0469 & 11.00  \\
J15562191+4757172 & 0.0195 & 10.49 & SE1 & 1.32 &16.33 & LCK-582705 & 0.0286 & 10.49  \\
J16024254+4111499 & 0.0333 & 11.11 & SS1 & 0.64 &12.57 & LCK-329416 & 0.0472 & 11.07  \\
J16024475+4111589 & 0.0333 & 10.78 & SS2 & 0.64 &12.57 & LCK-040350 & 0.0460 & 10.70  \\
J17045089+3448530 & 0.0568 & 11.01 & SS2 & 0.63 &7.76  & EN1-346329 & 0.0636 & 10.97  \\
J17045097+3449020 & 0.0568 & 11.28 & SS1 & 0.63 &7.76  & LCK-182514 & 0.0673 & 11.21\\
J20471908+0019150 & 0.0133 & 11.37 & SE1 & 0.99 &20.73 & LCK-515902 & 0.0723 & 11.37  \\
J13153076+6207447 & 0.0306 & 10.91 & SS2 & 1.34 &15.54 & LCK-347435 & 0.0468 & 10.87  \\
J13153506+6207287 & 0.0306 & 11.09 & SS1 & 1.34 &15.54 & LCK-048281 & 0.0481 & 11.05  \\
J09494143+0037163 & 0.0063 & 9.71 & SS2 & 2.04 &15.49 & NGC0024 & 0.0019 & 9.63  \\
J09495263+0037043 & 0.0063 & 9.95 & SS1 & 2.04 &15.49 & NGC2403 & 0.0004 & 9.99  \\
J13082737+0422125 & 0.0241 & 10.15 & SS2 & 1.29 &12.09 & NGC0925 & 0.0018 & 10.06  \\
J14530282+0317451 & 0.0052 & 9.92  & SS2 & 1.42 &10.09 & NGC3049 & 0.0050 & 9.91  \\
J14530523+0319541 & 0.0052 & 10.17 & SS1 & 1.42 &10.09 & NGC3184 & 0.0020 & 10.31  \\
\hline
\end{tabular}
\end{center}
\tablefoot{
\tablefoottext{1}{\scriptsize Category of the paired galaxies. The `SS' (or `SE') means the galaxy is in spiral-spiral (or spiral-elliptical) pairs. The `1' (or `2') means the galaxy is the primary (or secondary) component of the pair.}
\tablefoottext{2}{\scriptsize Normalized separation that calculated from Equation \ref{equ_sep}.}
}
\end{table*}

\subsection{Infrared data}
\label{subsec_irdata}

Results of \textit{Spitzer} IRAC and MIPS observations for the paired galaxies are provided by Xu10. The IR luminosity $L_{\rm IR}$ is calculated using the IRAC $8\,\mu$m and MIPS $24\,\mu$m data:
\begin{eqnarray}
\label{equ_calir}
\nonumber \log (L_{\rm IR} [L_{\odot}])&=&\log (L_{24}[L_{\odot}])+(0.87\pm0.03)\\
\nonumber &&+(0.56\pm0.09)\,\log (L_8/L_{24}),\\
{ }    
\end{eqnarray}
where $L_{\lambda}=\nu L_{\nu}$ at $\lambda$ $\mu$m. This estimation is consistent with the one given by \cite{dale_infrared_2005} using 24, 70, and 160 $\mu$m data\citep{xu_local_2010}. 

The IR data of control galaxies are selected from Spitzer SWIRE fields (34 galaxies) and Spitzer SINGS samples (5 galaxies). The fluxes of galaxies in SWIRE fields are taken from the SWIRE Data Release 2 \citep{surace2005}, and the fluxes of galaxies in SINGS sample are taken from \cite{smith_spitzer_2007}.

\subsection{Ultraviolet data}
\label{subsec_uvdata}

Ultraviolet images are taken from the GALEX GR6 database using GALEXVIEW. This database includes FUV and NUV images for 24 of our 27 pairs. These images are shown in Figure \ref{fig1}. Two methods are used to measure the UV fluxes:
\begin{enumerate}
  \item Manual photometry is applied using an IDL program. This program measures NUV and FUV fluxes for one source simultaneously, and the NUV images are used as the reference to detect sources. One starts the photometry with selecting a region by eye to confine the initial area where the photometry will be carried out. Then the program automatically searches for sources in this region and conducts aperture photometry using a set of elliptical apertures. The total flux density within the aperture is calculated where the growth curve converges. Last, the flux densities are corrected for Galactic extinction using a Schlegel map \citep{schlegel_maps_1998} and the Galactic extinction curve of \cite{cardelli_relationship_1989}. A detailed description of this method can be found in \cite{iglesias-paramo_star_2006}. This method has been proved reliable by several previous studies \citep[e.g.][]{iglesias-paramo_star_2006, buat_local_2007, takeuchi2012}.
  
  \item Classical aperture photometry is applied for pairs in which the components are extremely close to each other such that the manual program may not be able to rule out contamination from one component to the next when searching for the convergence radius (Figure \ref{fig1}). In addition, classical aperture photometry is also applied to source No.51 (upper right component of pair No.51-52). This galaxy is quite extended. As a result, the manual photometry defines a smaller aperture that covers only about half of the entire source.    
      
\end{enumerate}

\begin{figure}
\centering
\resizebox{\hsize}{!}{\includegraphics{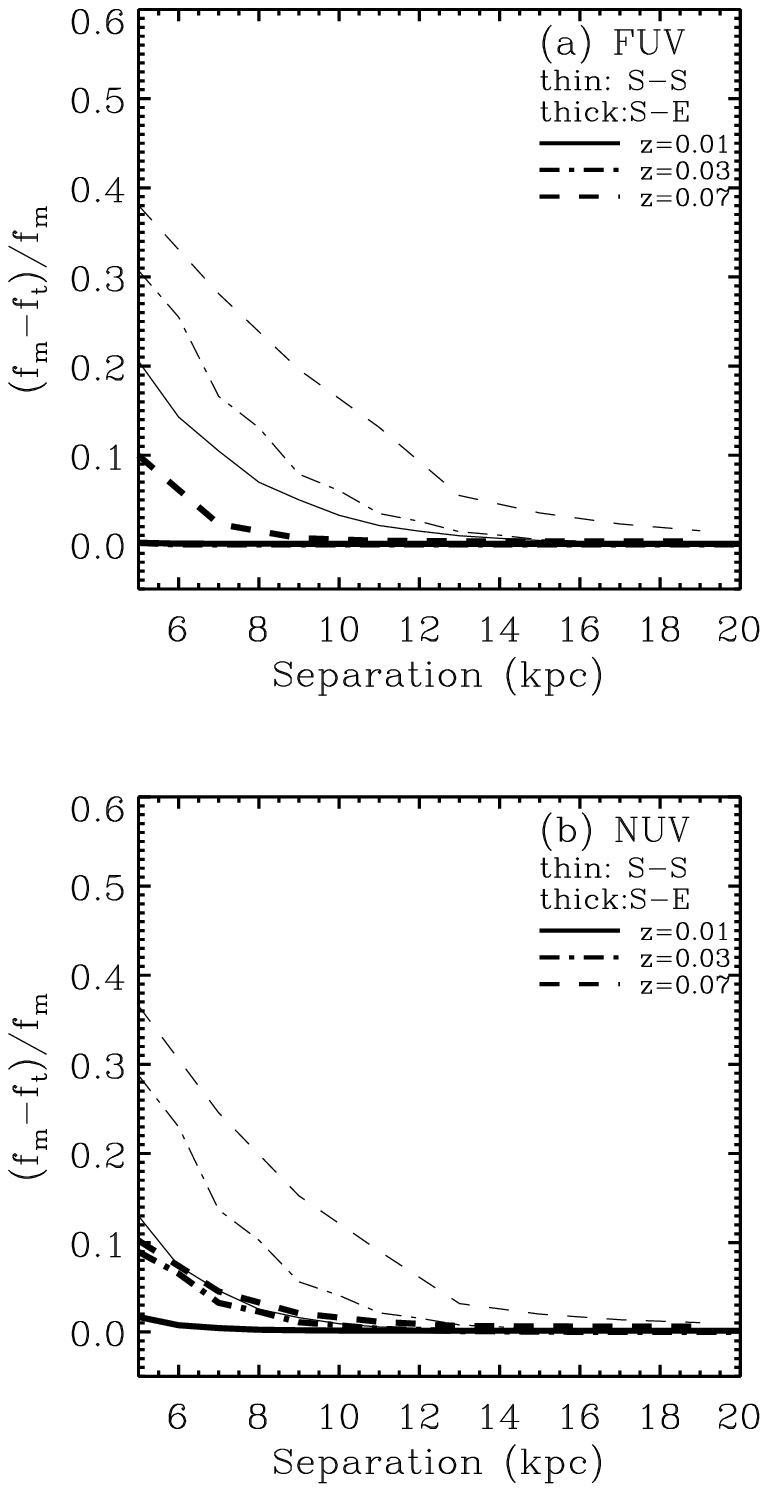}}
\caption{Contamination from one component to the next. The simulation uses stacked profiles of S and E galaxies in our pair sample. Results for S-S (thin lines) and S-E (thick lines) pairs at $z=0.01$ (solid), $z=0.03$ (dash dot), $z=0.07$ (dashed) are plotted.\label{fig_pho}}
\end{figure}

We simulated the contamination from one component to the next using stacked light profiles of our S and E galaxies. For round galaxies, we calculated the fluxes of one galaxy with a companion galaxy (measured flux, $f_{m}$) and without one (true flux, $f_{t}$) in fixed apertures. The value of $(f_{m}-f_{t})/f_{m}$ is shown in Figure \ref{fig_pho}. We find for most of our galaxies that the contamination is less than 5\%. For the pairs J0211-0039, the contamination is less than 10\%. For the pairs J1510+5810 and J1704+3448, the contamination may reach 20\%. On average, the contamination causes little error ($<$ 0.025 dex) in our result and does not affect our conclusion.

%___________________________________________________one-column figure

The results of the NUV and FUV photometry are shown in Table \ref{tb1}. The different methods of photometry applied to the sources are indicated as `M' and `A' for manual and aperture photometry, respectively. 

For the control sources, 38 sources are found in the GALEX image database. Manual photometry was applied and the results are shown in Table \ref{tb2}.

To prove the consistency between these methods, we compared the results of the different methods with GALEX pipeline data (Figure \ref{fig_comp_pho}). For paired galaxies, the results of manual photometry and aperture photometry are consistent with each other for sources that are separated well from their companion galaxies. Also, the results of both manual and aperture photometry show consistency with pipeline data when the sources are not too close to their companion galaxies and are not very extended. The GALEX pipeline photometry is not preferred because the automatic source extraction may identify extended sources as separate sources (referred to as {\it shredding} in \cite{takeuchi_star_2010}) or mistake a close pair as a single source, and therefore it is less accurate than our method.  

\subsection{Nuclear fluxes}
\label{subsec_nucdata}
We also measured fluxes within different circular apertures of each galaxy in order to examine the fraction of the integrated emission contributed by the nuclear and near-nuclear regions. For IR bands, IRAC fluxes within 4 kpc and 10 kpc apertures are already given by Xu10 for paired galaxies, so we only need to measure MIPS fluxes and the same apertures for the control galaxies using SWIRE and SINGS images. For UV bands, the fluxes within the same apertures were measured, with the goal of comparing the fraction of the nuclear contribution with the IR part. Tables \ref{tb1} and \ref{tb2} give the results for paired and control galaxies, respectively. The resolutions of GALEX images in the FUV and NUV bands are $\sim$ 4.3$''$ and $\sim$ 5.3$''$, while the resolutions of Spitzer images in the 8 and 24 $\mu$m bands are $\sim$ 2$''$ and $\sim$ 6$''$. For our galaxies, the 4 kpc central region is about $7''$ in size, and therefore the central regions of most of our galaxies can be resolved. For a few galaxies with small angular size, the difference in image resolutions may cause some uncertainties. However, it can be seen that the resolutions of GALEX and Spitzer MIPS are very similar. Although IRAC 8 $\mu$m has a higher resolution, the weight of 8 $\mu$m fluxes is very small when calculating the IR luminosity (Equation (\ref{equ_calir})). Therefore, the bias caused by the difference in resolutions can be ignored.

%%__________________________________tables in this section

\begin{table*}\scriptsize
\caption{GALEX NUV and FUV fluxes for paired galaxies.} \label{tb1}
\begin{center}
\begin{tabular}{rrccccccc}
\hline\hline
No. & Name & \multicolumn{3}{c}{NUV fluxes [$\mu$Jy]} & \multicolumn{3}{c}{FUV fluxes [$\mu$Jy]} & Pho\\                                             
 &   &Total &4kpc\tablefootmark{1} & 10kpc\tablefootmark{2} & Total  &4kpc\tablefootmark{1} & 10kpc\tablefootmark{2}  & \\
\hline

       1 & J00202580+0049350 & 135.86$\pm$1.81 & 79.99$\pm$1.10 & 116.66$\pm$1.55
 & 46.30$\pm$1.68 & 28.33$\pm$1.13 & 38.69$\pm$1.46 & M\\
       2 & J00202748+0050009 & 62.39$\pm$1.18 & 45.85$\pm$0.88 & 82.32$\pm$1.47 & 
31.26$\pm$1.26 & 25.74$\pm$1.08 & 38.63$\pm$1.48 & M\\
       3 & J01093371+0020322 & 14.99$\pm$0.77 & 2.34$\pm$0.21 & 8.89$\pm$0.44 & 
6.20$\pm$0.71 & 0.92$\pm$0.21 & 3.12$\pm$0.42 & M\\
       4 & J01093517+0020132 & 9.42$\pm$0.73 & 2.43$\pm$0.22 & 6.74$\pm$0.42 & 
3.16$\pm$0.62 & 1.04$\pm$0.23 & 2.62$\pm$0.40 & M\\
       5 & J01183417-0013416 & 41.88$\pm$1.31 & 12.25$\pm$0.57 & 36.28$\pm$1.13 & 
18.65$\pm$1.21 & 7.73$\pm$0.66 & 17.32$\pm$1.10 & M\\
       6 & J01183556-0013594 & 643.75$\pm$3.50 & 24.73$\pm$0.63 & 167.72$\pm$1.63
 & 423.20$\pm$4.60 & 15.01$\pm$0.84 & 108.57$\pm$2.27 & M\\
       7 & J02110638-0039191 & 92.12$\pm$1.90 & 17.10$\pm$0.76 & 62.61$\pm$1.57 & 
45.31$\pm$2.23 & 12.02$\pm$1.06 & 32.84$\pm$1.89 & M\\
       8 & J02110832-0039171 & 92.94$\pm$1.93 & 35.69$\pm$1.06 & 83.52$\pm$1.82 & 
10.63$\pm$1.44 & 5.37$\pm$0.77 & 10.84$\pm$1.34 & M\\
       9 & J09060283+5144411 & ... & ... & 
... & ... & ... & 
... & M\\
      10 & J09060498+5144071 & ... & ... & 
... & ... & ... & 
... & M\\
      11 & J09374413+0245394 & 1604.41$\pm$5.75 & 15.92$\pm$0.59 & 120.81$\pm$1.58
 & 990.88$\pm$9.45 & 8.05$\pm$0.87 & 65.12$\pm$2.44 & M\\
      12 & J09374506+0244504 & 58.24$\pm$4.40 & 25.61$\pm$1.38 & 52.39$\pm$3.24 & 
23.51$\pm$3.43 & 12.91$\pm$1.37 & 23.86$\pm$2.68 & M\\
      13 & J09494143+0037163 & 3654.47$\pm$11.51 & 2484.03$\pm$8.70 & 
3764.10$\pm$11.83 & 2406.68$\pm$14.94 & 1678.75$\pm$12.15 & 2432.11$\pm$15.06 & 
M\\
      14 & J09495263+0037043 & 8222.12$\pm$16.51 & 1999.33$\pm$7.56 & 
6423.39$\pm$13.79 & 6051.09$\pm$23.63 & 1388.66$\pm$10.92 & 4690.21$\pm$20.23 & 
M\\
      15 & J10205188+4831096 & 46.47$\pm$5.72 & 3.31$\pm$1.08 & 17.78$\pm$2.48 & 
26.10$\pm$5.95 & 4.35$\pm$2.03 & 11.00$\pm$3.27 & M\\
      16 & J10205369+4831246 & 17.51$\pm$3.00 & 0.93$\pm$0.67 & 4.65$\pm$1.54 & 
10.61$\pm$3.49 & 0.77$\pm$0.91 & 1.95$\pm$1.57 & M\\
      17 & J10272950+0114490 & 196.57$\pm$2.94 & 68.00$\pm$1.40 & 148.24$\pm$2.22
 & 82.45$\pm$3.10 & 34.35$\pm$1.71 & 67.14$\pm$2.51 & M\\
      18 & J10272970+0115170 & 33.17$\pm$3.04 & 16.43$\pm$2.20 & 29.29$\pm$5.34 & 
7.52$\pm$1.12 & 4.15$\pm$0.70 & 7.44$\pm$1.25 & M\\
      19 & J10435053+0645466 & ... & ... & 
... & ... & ... & 
... & M\\
      20 & J10435268+0645256 & ... & ... & 
... & ... & ... & 
... & M\\
      21 & J10514368+5101195 & 159.89$\pm$9.03 & 30.16$\pm$3.38 & 82.13$\pm$5.83
 & 58.98$\pm$9.58 & 15.68$\pm$4.26 & 35.80$\pm$6.68 & M\\
      22 & J10514450+5101303 & 10.19$\pm$2.04 & 16.47$\pm$2.62 & 87.00$\pm$6.10 & 
6.08$\pm$2.71 & 4.14$\pm$2.40 & 40.69$\pm$7.24 & M\\
      23 & J12020424+5342317 & ... & ... & 
... & ... & ... & 
... & M\\
      24 & J12020537+5342487 & ... & ... & 
... & ... & ... & 
... & M\\
      25 & J13082737+0422125 & 220.22$\pm$9.96 & 53.73$\pm$4.56 & 177.91$\pm$8.53
 & 147.73$\pm$13.59 & 38.83$\pm$6.72 & 120.69$\pm$12.01 & M\\
      26 & J13082964+0422045 & 389.70$\pm$13.47 & 81.05$\pm$5.56 & 238.50$\pm$9.76
 & 356.09$\pm$20.83 & 54.29$\pm$7.93 & 205.57$\pm$15.49 & M\\
      27 & J13325525-0301347 & 16.53$\pm$2.16 & 13.77$\pm$1.97 & 51.32$\pm$3.90 & 
13.46$\pm$3.27 & 11.05$\pm$2.97 & 23.42$\pm$4.44 & M\\
      28 & J13325655-0301395 & 537.81$\pm$12.13 & 18.67$\pm$2.19 & 164.06$\pm$6.50
 & 332.13$\pm$16.30 & 7.15$\pm$2.37 & 104.20$\pm$9.01 & M\\
      29 & J13462001-0325407 & 163.96$\pm$10.50 & 12.73$\pm$2.81 & 84.38$\pm$7.15
 & 104.15$\pm$13.43 & 9.25$\pm$3.87 & 56.96$\pm$9.59 & M\\
      30 & J13462215-0325057 & 30.84$\pm$6.16 & 14.20$\pm$2.98 & 23.04$\pm$5.18 & 
9.79$\pm$6.30 & ... & 10.59$\pm$5.31 & M\\
      31 & J14005782+4251207 & 71.52$\pm$1.74 & 15.36$\pm$0.61 & 44.64$\pm$1.10 & 
38.37$\pm$2.46 & 9.21$\pm$0.97 & 26.23$\pm$1.70 & M\\
      32 & J14005882+4250427 & 94.74$\pm$1.78 & 26.28$\pm$0.78 & 72.12$\pm$1.33 & 
49.99$\pm$2.59 & 16.39$\pm$1.27 & 38.00$\pm$1.99 & M\\
      33 & J14250552+0313590 & 62.01$\pm$1.68 & 15.34$\pm$0.67 & 49.48$\pm$1.30 & 
28.97$\pm$1.85 & 9.27$\pm$0.89 & 25.22$\pm$1.55 & M\\
      34 & J14250739+0313560 & 15.87$\pm$1.13 & 5.40$\pm$0.47 & 13.57$\pm$0.93 & 
5.58$\pm$1.12 & 3.07$\pm$0.56 & 5.16$\pm$0.93 & M\\
      35 & J14334683+4004512 & 387.57$\pm$13.72 & 15.51$\pm$2.46 & 98.20$\pm$6.17
 & 219.93$\pm$17.53 & 7.97$\pm$3.09 & 46.35$\pm$7.47 & M\\
      36 & J14334840+4005392 & 509.31$\pm$14.25 & 149.34$\pm$7.17 & 
361.48$\pm$11.28 & 300.64$\pm$19.05 & 66.99$\pm$8.54 & 214.30$\pm$15.39 & M\\
      37 & J14530282+0317451 & 1691.30$\pm$6.15 & 1021.81$\pm$3.68 & 
1689.23$\pm$6.35 & 1039.03$\pm$8.30 & 626.87$\pm$6.09 & 1024.08$\pm$8.36 & M\\
      38 & J14530523+0319541 & 2562.27$\pm$6.82 & 436.93$\pm$2.36 & 
2088.87$\pm$5.35 & 1667.60$\pm$11.10 & 220.25$\pm$3.77 & 1334.07$\pm$9.31 & M\\
      39 & J15064391+0346364 & 2.17$\pm$0.45 & 7.16$\pm$0.60 & 21.26$\pm$1.21 & 
0.81$\pm$0.44 & 2.68$\pm$0.61 & 5.21$\pm$1.08 & M\\
      40 & J15064579+0346214 & 294.63$\pm$3.68 & 12.89$\pm$0.66 & 60.87$\pm$1.46
 & 195.30$\pm$4.67 & 7.43$\pm$0.85 & 35.09$\pm$1.86 & M\\
      41 & J15101587+5810425 & 6.46$\pm$1.25 & 8.11$\pm$1.55 & 33.97$\pm$3.55 & 
1.93$\pm$1.08 & 2.84$\pm$1.32 & 8.93$\pm$2.53 & A\\
      42 & J15101776+5810375 & 88.00$\pm$4.23 & 25.07$\pm$2.09 & 69.59$\pm$3.60 & 
50.70$\pm$5.43 & 19.43$\pm$3.21 & 40.35$\pm$4.72 & A\\
      43 & J15281276+4255474 & 261.10$\pm$8.47 & 61.55$\pm$3.70 & 199.60$\pm$6.99
 & 105.60$\pm$9.32 & 18.70$\pm$3.60 & 73.22$\pm$7.36 & M\\
      44 & J15281667+4256384 & 47.82$\pm$4.83 & 24.77$\pm$2.51 & 43.78$\pm$4.18 & 
11.86$\pm$4.58 & 7.11$\pm$2.42 & 14.53$\pm$4.24 & M\\
      45 & J15562191+4757172 & 155.66$\pm$4.45 & 38.89$\pm$1.95 & 121.53$\pm$3.62
 & 77.24$\pm$9.38 & 21.54$\pm$4.44 & 67.72$\pm$8.15 & M\\
      46 & J15562738+4757302 & $<$ 1.45  & ... & ...
 & $<$ 0.91 & ... & ... & M\\
      47 & J16024254+4111499 & 932.13$\pm$10.04 & 91.58$\pm$3.11 & 376.17$\pm$6.33
 & 607.65$\pm$22.66 & 66.52$\pm$7.42 & 253.73$\pm$14.54 & M\\
      48 & J16024475+4111589 & 345.27$\pm$6.38 & 55.92$\pm$2.48 & 231.41$\pm$5.15
 & 197.24$\pm$13.01 & 39.18$\pm$5.71 & 138.59$\pm$10.85 & M\\
      49 & J17045089+3448530 & 14.98$\pm$2.02 & 1.59$\pm$0.73 & 30.36$\pm$2.86 & 
8.81$\pm$2.65 & 2.72$\pm$1.47 & 13.36$\pm$3.29 & A\\
      50 & J17045097+3449020 & 141.08$\pm$6.07 & 23.12$\pm$2.37 & 89.97$\pm$4.71
 & 54.15$\pm$6.57 & 9.34$\pm$2.66 & 41.64$\pm$5.64 & A\\
      51 & J20471908+0019150 & 2906.97$\pm$15.56 & 20.55$\pm$1.18 & 
165.51$\pm$3.01 & 1644.18$\pm$14.13 & 11.75$\pm$1.10 & 71.46$\pm$2.72 & A\\
      52 & J20472428+0018030 & 114.65$\pm$3.03 & 66.15$\pm$1.94 & 124.30$\pm$4.51
 & 31.58$\pm$2.28 & 18.95$\pm$1.34 & 35.59$\pm$2.61 & M\\
      53 & J13153076+6207447 & 329.93$\pm$6.00 & 30.69$\pm$1.86 & 171.58$\pm$4.47
 & 191.07$\pm$15.65 & 12.72$\pm$4.05 & 89.59$\pm$10.73 & A\\
      54 & J13153506+6207287 & 736.30$\pm$9.99 & 130.55$\pm$3.65 & 367.29$\pm$6.24
 & 399.30$\pm$23.76 & 80.85$\pm$9.95 & 198.45$\pm$15.68 & A\\
\hline
\end{tabular}
\end{center}
\tablefoot{
\tablefoottext{1}{\scriptsize Fluxes inside $4$ kpc aperture.}
\tablefoottext{2}{\scriptsize Fluxes inside $10$ kpc aperture.}
}
\end{table*}

\begin{table*}\scriptsize
%\centering
\begin{center}
\caption{GALEX NUV and FUV fluxes for control galaxies.}
\label{tb2}
\begin{tabular}{rrccccccc}
\hline\hline
No.& ID & \multicolumn{3}{c}{NUV fluxes ($\mu$Jy)} & \multicolumn{3}{c}{FUV fluxes ($\mu$Jy)} & Paired galaxy ID\\   
 & & Total & 4 kpc\tablefootmark{1} & 10 kpc\tablefootmark{2} & Total & 4 kpc\tablefootmark{1} & 10 kpc\tablefootmark{2} & \\                                      
\hline

       1 & LCK-287434 & 26.69$\pm$0.59 & 4.28$\pm$0.13 & 13.71$\pm$0.28 & 
13.47$\pm$0.52 & 2.16$\pm$0.14 & 6.46$\pm$0.27 & J00202580+0049350\\
       2 & LCK-178064 & 32.52$\pm$0.48 & 3.06$\pm$0.09 & 14.58$\pm$0.21 & 
17.06$\pm$0.44 & 1.53$\pm$0.10 & 7.89$\pm$0.24 & J00202748+0050009\\
       3 & LCK-320371 & 89.64$\pm$4.18 & 8.12$\pm$0.47 & 39.81$\pm$1.22 & 
64.74$\pm$0.74 & 6.07$\pm$0.20 & 28.30$\pm$0.43 & J01093371+0020322\\
       4 & LCK-523686 & 20.38$\pm$0.41 & 2.67$\pm$0.07 & 8.70$\pm$0.14 & 
6.34$\pm$0.39 & 0.78$\pm$0.08 & 2.52$\pm$0.16 & J01093517+0020132\\
       5 & LCK-415950 & 1147.61$\pm$1.29 & 71.78$\pm$0.30 & 279.07$\pm$0.59 & 
671.83$\pm$2.25 & 39.00$\pm$0.53 & 141.13$\pm$1.00 & J01183417-0013416\\
       6 & LCK-086596 & 130.89$\pm$0.49 & 4.14$\pm$0.07 & 19.25$\pm$0.15 & 
97.10$\pm$0.72 & 2.87$\pm$0.12 & 13.71$\pm$0.26 & J01183556-0013594\\
       7 & EN1-158103 & 423.57$\pm$1.20 & 48.20$\pm$0.37 & 183.86$\pm$0.74 & 
298.52$\pm$2.32 & 30.50$\pm$0.72 & 116.71$\pm$1.41 & J02110638-0039191\\
       8 & EN1-360222 & 85.84$\pm$0.59 & 6.52$\pm$0.13 & 37.18$\pm$0.31 & 
44.99$\pm$1.31 & 4.59$\pm$0.35 & 23.99$\pm$0.82 & J02110832-0039171\\
       9 & EN1-010947 & 8.43$\pm$0.26 & 2.15$\pm$0.08 & 7.41$\pm$0.19 & 
2.25$\pm$0.32 & 0.69$\pm$0.14 & 1.99$\pm$0.28 & J09060283+5144411\\
      10 & LCK-162208 & 199.59$\pm$0.83 & 36.47$\pm$0.29 & 136.22$\pm$0.59 & 
98.87$\pm$0.86 & 18.38$\pm$0.35 & 73.85$\pm$0.71 & J09060498+5144071\\
      11 & EN1-018834 & 10.43$\pm$0.51 & 1.60$\pm$0.07 & 4.71$\pm$0.13 & 
3.94$\pm$0.59 & 1.10$\pm$0.15 & 2.64$\pm$0.24 & J09374413+0245394\\
      12 & LCK-233199 & 62.65$\pm$0.75 & 13.15$\pm$0.23 & 45.42$\pm$0.47 & 
40.17$\pm$0.81 & 9.67$\pm$0.33 & 29.63$\pm$0.60 & J09374506+0244504\\
      13 & LCK-019297 & 15.80$\pm$0.65 & 2.61$\pm$0.10 & 10.31$\pm$0.24 & 
5.99$\pm$0.51 & 1.22$\pm$0.11 & 4.11$\pm$0.23 & J09494143+0037163\\
      14 & LCK-703238 & 52.11$\pm$0.58 & 3.39$\pm$0.07 & 18.96$\pm$0.19 & 
35.53$\pm$0.63 & 2.61$\pm$0.14 & 13.48$\pm$0.32 & J09495263+0037043\\
      15 & LCK-050667 & 39.92$\pm$0.64 & 3.62$\pm$0.13 & 18.22$\pm$0.30 & 
22.90$\pm$0.65 & 1.94$\pm$0.15 & 10.86$\pm$0.37 & J10205188+4831096\\
      16 & LCK-027930 & 59.00$\pm$0.60 & 5.04$\pm$0.08 & 25.83$\pm$0.20 & 
29.76$\pm$0.64 & 2.71$\pm$0.14 & 14.10$\pm$0.33 & J10205369+4831246\\
      17 & LCK-071868 & 297.51$\pm$0.48 & 26.51$\pm$0.11 & 132.75$\pm$0.26 & 
185.34$\pm$0.92 & 16.75$\pm$0.26 & 84.19$\pm$0.59 & J10272950+0114490\\
      18 & EN1-516050 & ... & ... & 
... & ... & ... & 
... & J10272970+0115170\\
      19 & LCK-641925 & 461.08$\pm$0.99 & 13.38$\pm$0.14 & 92.53$\pm$0.37 & 
301.58$\pm$1.51 & 7.52$\pm$0.23 & 53.62$\pm$0.60 & J10435053+0645466\\
      20 & LCK-400414 & 105.03$\pm$0.85 & 11.42$\pm$0.09 & 38.10$\pm$0.19 & 
69.68$\pm$0.96 & 5.80$\pm$0.16 & 22.01$\pm$0.33 & J10435268+0645256\\
      21 & LCK-534543 & 306.47$\pm$0.93 & 11.40$\pm$0.14 & 56.46$\pm$0.32 & 
199.36$\pm$1.31 & 6.20$\pm$0.22 & 32.88$\pm$0.50 & J10514368+5101195\\
      22 & LCK-136060 & 26.52$\pm$0.87 & 2.21$\pm$0.12 & 12.18$\pm$0.31 & 
14.59$\pm$0.63 & 1.09$\pm$0.11 & 7.37$\pm$0.29 & J10514450+5101303\\
      23 & LCK-172179 & 36.33$\pm$0.51 & 1.93$\pm$0.06 & 11.58$\pm$0.17 & 
18.48$\pm$0.52 & 1.16$\pm$0.10 & 6.76$\pm$0.24 & J12020424+5342317\\
      24 & LCK-564807 & 132.62$\pm$0.54 & 5.93$\pm$0.08 & 39.19$\pm$0.21 & 
94.23$\pm$0.86 & 3.88$\pm$0.16 & 25.67$\pm$0.41 & J12020537+5342487\\
      25 & LCK-621286 & 15.13$\pm$0.44 & 2.80$\pm$0.08 & 9.63$\pm$0.18 & 
5.74$\pm$0.37 & 1.27$\pm$0.10 & 4.14$\pm$0.19 & J13082737+0422125\\
      26 & LCK-038716 & 96.19$\pm$0.66 & 6.35$\pm$0.14 & 31.60$\pm$0.31 & 
65.93$\pm$0.77 & 4.17$\pm$0.18 & 22.14$\pm$0.42 & J13082964+0422045\\
      27 & LCK-582705 & 23.95$\pm$0.37 & 6.74$\pm$0.10 & 18.54$\pm$0.19 & 
7.45$\pm$0.39 & 2.66$\pm$0.14 & 6.30$\pm$0.24 & J13325525-0301347\\
      28 & LCK-329416 & 35.87$\pm$0.48 & 3.67$\pm$0.08 & 19.44$\pm$0.20 & 
17.83$\pm$0.53 & 1.93$\pm$0.12 & 9.67$\pm$0.28 & J13325655-0301395\\
      29 & LCK-040350 & 2.47$\pm$0.24 & 0.48$\pm$0.08 & 1.93$\pm$0.18 & 
0.69$\pm$0.20 & 0.16$\pm$0.06 & 0.41$\pm$0.14 & J13462001-0325407\\
      30 & EN1-346329 & 80.04$\pm$0.66 & 5.25$\pm$0.13 & 27.97$\pm$0.29 & 
49.61$\pm$1.09 & 3.25$\pm$0.24 & 18.06$\pm$0.56 & J13462215-0325057\\
      31 & LCK-182514 & 50.07$\pm$0.62 & 1.99$\pm$0.07 & 9.22$\pm$0.16 & 
32.31$\pm$0.61 & 1.39$\pm$0.10 & 5.89$\pm$0.20 & J14005782+4251207\\
      32 & LCK-515902 & 9.38$\pm$0.44 & 0.53$\pm$0.04 & 2.99$\pm$0.09 & 
3.16$\pm$0.37 & 0.26$\pm$0.05 & 1.14$\pm$0.11 & J14005882+4250427\\
      33 & LCK-347435 & 8.74$\pm$0.35 & 0.88$\pm$0.05 & 3.32$\pm$0.11 & 
2.50$\pm$0.39 & 0.36$\pm$0.07 & 1.05$\pm$0.13 & J14250552+0313590\\
      34 & LCK-048281 & 32.83$\pm$0.48 & 1.33$\pm$0.06 & 6.70$\pm$0.13 & 
18.38$\pm$0.48 & 0.53$\pm$0.06 & 2.97$\pm$0.15 & J14250739+0313560\\
      35 & NGC0024 & 11029.62$\pm$19.40 & 6024.89$\pm$12.83 & 10153.03$\pm$18.43
 & 8081.00$\pm$26.50 & 4344.08$\pm$18.78 & 7450.55$\pm$25.41 & J14334683+4004512
\\
      36 & NGC2403 & 317739.62$\pm$76.09 & 234299.82$\pm$60.46 & 
326425.75$\pm$84.24 & 234466.58$\pm$105.82 & 166958.51$\pm$87.71 & 
241057.21$\pm$111.09 & J14334840+4005392\\
      37 & NGC0925 & 65670.66$\pm$62.90 & 12313.23$\pm$22.70 & 
38203.55$\pm$41.90 & 48593.78$\pm$84.14 & 8666.34$\pm$32.27 & 27602.99$\pm$59.02
 & J14530282+0317451\\
      38 & NGC3049 & 4639.48$\pm$9.40 & 2575.06$\pm$6.29 & 4168.10$\pm$8.61 & 
2872.98$\pm$17.44 & 1506.02$\pm$12.19 & 2509.49$\pm$16.10 & J14530523+0319541\\
      39 & NGC3184 & 47886.38$\pm$153.04 & 5218.17$\pm$45.98 & 
30169.53$\pm$110.95 & 34313.52$\pm$212.38 & 3101.25$\pm$61.80 & 
20133.21$\pm$157.35 & J15064391+0346364\\
\hline

\end{tabular}
\end{center}
\tablefoot{
\tablefoottext{1}{\scriptsize Fluxes inside $4$ kpc aperture.}
\tablefoottext{2}{\scriptsize Fluxes inside $10$ kpc aperture.}
}
\end{table*}

%%%%_________________________________

\section{Dust attenuation}
\label{sec_dustatt}

\begin{table*}\scriptsize
\setlength{\tabcolsep}{0.02in}
\centering
\caption{SFR, SSFR, and $A_{\rm FUV}$ for non-AGN spirals in the pair and control samples.} \label{tb3}
\begin{tabular}{cccccccccccccc}
\hline\hline
Paired Galaxy ID & ${\rm SFR}_{\rm TOT}$ & $\log$SFR$_{\rm FUV}$& $\log$SFR$_{\rm IR}$ & $\log$SSFR$_{\rm TOT}$ & $\frac{\rm SFR_{FUV}}{\rm SFR_{TOT}}$ & $A_{\rm FUV}$ & Control Galaxy ID & ${\rm SFR}_{\rm TOT}$ & $\log$SFR$_{\rm FUV}$ & $\log$SFR$_{\rm IR}$ &$\log$SSFR$_{\rm TOT}$ & $\frac{\rm SFR_{FUV}}{\rm SFR_{TOT}}$ & $A_{\rm FUV}$ \\ 
 & ($M_{\odot}$yr$^{-1}$) & ($M_{\odot}$yr$^{-1}$) & ($M_{\odot}$yr$^{-1}$) & (yr$^{-1}$) &  & (mag) &  & ($M_{\odot}$yr$^{-1}$) & ($M_{\odot}$yr$^{-1}$)& ($M_{\odot}$yr$^{-1}$) & (yr$^{-1}$) &  & (mag)\\
\hline 

J00202580+0049350 & 0.13 & -1.39 & 0.27 & -10.71 & 3.1\% & 4.06 & LCK-287434 & -0.06 & -1.36 & 0.07 & -10.94 & 5.0\% & 3.47 \\
J01093517+0020132 & -0.91 & -1.70 & -0.83 & -11.96 & 16.3\% & 2.10 & LCK-178064 & -0.16 & -0.95 & -0.08 & -11.31 & 16.1\% & 2.12 \\
J01183556-0013594 & 0.94 & 0.49 & 0.90 & -9.99 & 35.8\% & 1.24 & LCK-320371 & 0.18 & -0.33 & 0.18 & -10.78 & 30.5\% & 1.42 \\
J02110832-0039171 & -0.76 & -1.88 & -0.64 & -11.74 & 7.7\% & 2.98 & LCK-523686 & -0.53 & -1.38 & -0.44 & -11.47 & 14.1\% & 2.27 \\
J09374413+0245394 & 0.90 & 0.23 & 0.95 & -10.56 & 21.2\% & 1.81 & LCK-415950 & 0.87 & 0.33 & 0.88 & -10.50 & 28.7\% & 1.48 \\
J10205188+4831096 & 0.53 & -0.62 & 0.65 & -10.35 & 7.1\% & 3.07 & LCK-086596 & 0.11 & -0.16 & -0.07 & -10.75 & 53.8\% & 0.81 \\
J10272950+0114490 & 0.43 & -0.89 & 0.56 & -10.30 & 4.8\% & 3.52 & EN1-158103 & 0.47 & -0.08 & 0.48 & -10.23 & 28.4\% & 1.49 \\
J10435053+0645466 & ... & ... & 1.02 & ... & ... & ... & EN1-360222 & 0.52 & -0.57 & 0.64 & -10.22 & 8.0\% & 2.92 \\
J10435268+0645256 & ... & ... & 0.06 & ... & ... & ... & EN1-010947 & -1.32 & -2.01 & -1.26 & -11.99 & 20.1\% & 1.87 \\
J10514450+5101303 & -0.97 & -1.97 & -0.86 & -12.10 & 10.0\% & 2.66 & LCK-162208 & 0.56 & -0.75 & 0.69 & -10.56 & 4.9\% & 3.50 \\
J12020424+5342317 & ... & ... & 0.41 & ... & ... & ... & EN1-018834 & -0.03 & -1.28 & 0.10 & -11.09 & 5.6\% & 3.36 \\
J13082964+0422045 & 0.02 & -0.19 & -0.24 & -10.51 & 61.8\% & 0.66 & LCK-233199 & -0.15 & -1.04 & -0.05 & -10.75 & 12.8\% & 2.38 \\
J13325525-0301347 & 0.72 & -1.00 & 0.87 & -10.18 & 1.9\% & 4.60 & LCK-019297 & 0.92 & -1.37 & 1.07 & -10.04 & 0.5\% & 5.64 \\
J13325655-0301395 & 0.80 & 0.37 & 0.75 & -10.41 & 37.3\% & 1.20 & LCK-703238 & -0.04 & -0.65 & -0.01 & -11.24 & 24.8\% & 1.64 \\
J13462001-0325407 & -0.04 & -0.76 & 0.02 & -11.05 & 19.2\% & 1.92 & LCK-050667 & -0.23 & -0.81 & -0.21 & -11.15 & 26.4\% & 1.57 \\
J14005782+4251207 & 0.88 & -0.90 & 1.03 & -10.13 & 1.7\% & 4.72 & LCK-027930 & 0.50 & -0.70 & 0.63 & -10.56 & 6.3\% & 3.20 \\
J14005882+4250427 & 1.05 & -0.76 & 1.20 & -9.85 & 1.5\% & 4.80 & LCK-071868 & 0.84 & 0.11 & 0.90 & -10.10 & 19.0\% & 1.93 \\
J14250739+0313560 & -0.86 & -1.64 & -0.79 & -11.52 & 16.9\% & 2.06 & EN1-516050 & ... & ... & 0.28 & ... & ... & ... \\
J14334683+4004512 & 0.68 & -0.35 & 0.79 & -10.57 & 9.5\% & 2.73 & LCK-641925 & -0.09 & -0.15 & -0.78 & -11.28 & 85.8\% & 0.21 \\
J14334840+4005392 & 0.90 & -0.20 & 1.02 & -10.20 & 8.0\% & 2.93 & LCK-400414 & -0.05 & -0.76 & 0.01 & -11.17 & 19.5\% & 1.90 \\
J15064391+0346364 & -0.47 & -2.50 & -0.32 & -11.69 & 0.9\% & 5.26 & LCK-534543 & 0.57 & -0.21 & 0.65 & -10.57 & 16.6\% & 2.08 \\
J15064579+0346214 & 0.71 & -0.15 & 0.80 & -10.46 & 13.9\% & 2.28 & LCK-136060 & -0.23 & -0.96 & -0.16 & -11.31 & 18.6\% & 1.96 \\
J15101587+5810425 & -0.43 & -2.24 & -0.28 & -11.45 & 1.5\% & 4.81 & LCK-172179 & -0.18 & -0.90 & -0.12 & -11.20 & 19.3\% & 1.91 \\
J15101776+5810375 & 0.39 & -0.79 & 0.52 & -10.40 & 6.6\% & 3.16 & LCK-564807 & 0.30 & -0.22 & 0.30 & -10.46 & 30.2\% & 1.43 \\
J15281276+4255474 & 0.50 & -0.94 & 0.64 & -10.76 & 3.6\% & 3.87 & LCK-621286 & -0.45 & -1.42 & -0.35 & -11.77 & 10.9\% & 2.57 \\
J15281667+4256384 & -0.67 & -1.93 & -0.54 & -11.70 & 5.5\% & 3.38 & LCK-038716 & 0.18 & -0.33 & 0.18 & -10.82 & 30.7\% & 1.41 \\
J15562191+4757172 & 0.13 & -1.06 & 0.26 & -10.36 & 6.5\% & 3.18 & LCK-582705 & -0.91 & -1.72 & -0.83 & -11.40 & 15.6\% & 2.15 \\
J16024254+4111499 & 1.11 & 0.33 & 1.19 & -10.00 & 16.6\% & 2.09 & LCK-329416 & 0.25 & -0.89 & 0.37 & -10.82 & 7.3\% & 3.04 \\
J16024475+4111589 & 0.55 & -0.16 & 0.61 & -10.23 & 19.4\% & 1.91 & LCK-040350 & -0.34 & -2.33 & -0.19 & -11.04 & 1.0\% & 5.17 \\
J17045089+3448530 & 0.88 & -1.02 & 1.03 & -10.13 & 1.2\% & 5.00 & EN1-346329 & 0.43 & -0.18 & 0.46 & -10.54 & 24.8\% & 1.64 \\
J17045097+3449020 & 1.44 & -0.25 & 1.59 & -9.84 & 2.0\% & 4.51 & LCK-182514 & 0.55 & -0.31 & 0.64 & -10.66 & 13.8\% & 2.30 \\
J20471908+0019150 & 0.49 & -0.00 & 0.47 & -10.88 & 32.6\% & 1.35 & LCK-515902 & 0.19 & -1.26 & 0.33 & -11.18 & 3.6\% & 3.88 \\
J13153076+6207447 & 1.10 & -0.25 & 1.24 & -9.81 & 4.4\% & 3.62 & LCK-347435 & 0.18 & -1.75 & 0.33 & -10.69 & 1.2\% & 5.06 \\
J13153506+6207287 & 1.73 & 0.07 & 1.88 & -9.36 & 2.2\% & 4.44 & LCK-048281 & -0.20 & -0.86 & -0.15 & -11.25 & 21.8\% & 1.78 \\
J09494143+0037163 & -0.32 & -0.55 & -0.54 & -10.03 & 58.3\% & 0.72 & NGC0024 & -0.84 & -1.05 & -1.09 & -10.47 & 60.8\% & 0.68 \\
J09495263+0037043 & 0.09 & -0.12 & -0.16 & -9.86 & 60.9\% & 0.68 & NGC2403 & -0.30 & -0.95 & -0.26 & -10.29 & 22.7\% & 1.73 \\
J13082737+0422125 & -0.20 & -0.58 & -0.29 & -10.35 & 42.5\% & 1.06 & NGC0925 & -0.04 & -0.32 & -0.21 & -10.10 & 52.5\% & 0.84 \\
J14530282+0317451 & -0.63 & -1.06 & -0.67 & -10.55 & 36.8\% & 1.22 & NGC3049 & -0.11 & -0.66 & -0.10 & -10.02 & 28.2\% & 1.50 \\
J14530523+0319541 & -0.43 & -0.86 & -0.48 & -10.60 & 37.5\% & 1.19 & NGC3184 & 0.04 & -0.38 & -0.02 & -10.27 & 38.3\% & 1.17 \\
\hline
\end{tabular}
\end{table*}

As mentioned in Section \ref{sec_intro}, dust attenuation in paired galaxies may be very complex and is usually studied using numerical simulations. Here with both UV and IR data, we can examine whether there is a statistical difference in dust attenuation between pair and isolated galaxies. The attenuation in the FUV band, $A_{\rm FUV}$, of these galaxies can be calculated using the formula given in \cite{buat_spectral_2011}:
\begin{eqnarray}
\nonumber A_{\rm FUV}\textrm{[mag]} &=& 0.483 + 0.812\ y + 0.373\ y^2 + 0.299\ y^3\\
&&- 0.106\ y^4,
\label{equ_ext}
\end{eqnarray}
where $y\ =\ \log(L_{\rm IR}/L_{\rm FUV})$, $L_{\rm IR}$ is defined in Equation (\ref{equ_calir}), and $L_{\rm FUV}$ is $\nu L_{\nu}$ at the FUV band (1530 \AA).

The results for the 39 non-AGN spirals\footnote{Hereafter, we omit the term `non-AGN' since in the following analyses we only deal with these non-AGN spiral galaxies.} and the control sample are presented in Table \ref{tb3}. There are several different attenuation measurements in the literature. Although these measurements depend on star formation history, it has been shown that they deviate little from each other: \cite{buat_spectral_2011} compared their results with \cite{meurer_dust_1999} and \cite{buat_dust_2005}, and show that the difference between them is at most 0.3 magnitude. The difference between their results and those of \cite{cortese_ultraviolet_2008} is small for low dust attenuation, but can reach 0.7 mag for high attenuations. \cite{boquien_irx-_2012} also used a fourth-order polynomial to fit the relation, and they show that the difference between their work and others \citep{burgarella_ultraviolet--far_2006, cortese_ultraviolet_2008, buat_spectral_2011, hao_dust-corrected_2011} is smaller than 0.2 magnitude. To assure that using different formulae does not affect our conclusion, we also use \cite{buat_dust_2005} and \cite{cortese_ultraviolet_2008} to calculate the dust attenuation, and find our conclusion unchanged.  

\begin{figure}
\centering
\resizebox{\hsize}{!}{\includegraphics{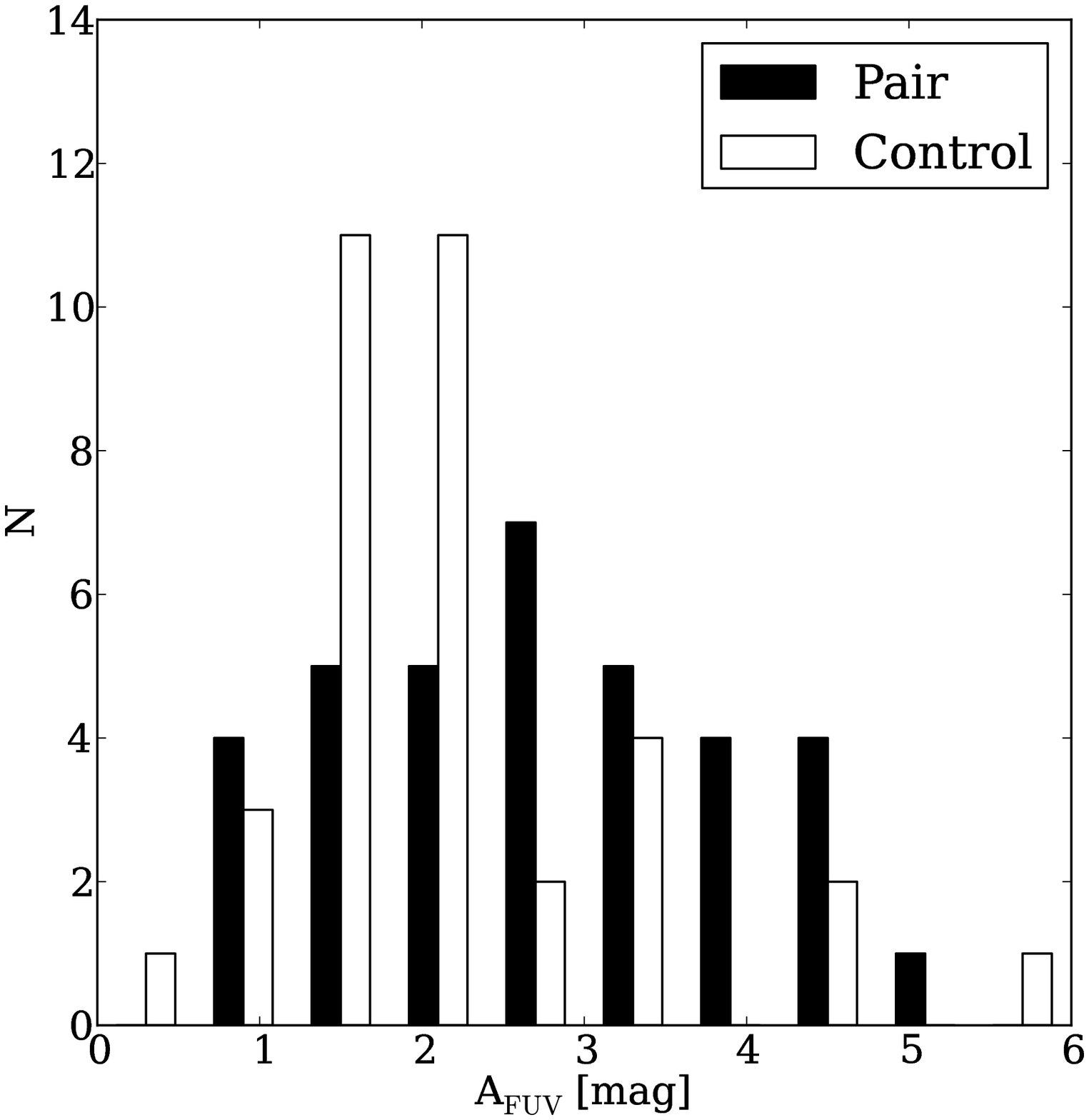}}
\caption{Histograms of $A_{\rm FUV}$ distributions for spirals in the pair and control samples. \label{fig_ext1}}
\end{figure}

Figure \ref{fig_ext1} plots the histogram of the distribution of dust attenuation in our pair and control samples. It can be seen that $A_{\rm FUV}$ in paired galaxies has a very different distribution than in the control sample. Quantitatively, the mean $A_{\rm FUV}$ is $2.82\pm0.24$ mag for spirals in the pair sample, whereas it is only $2.20\pm0.21$ mag for the control galaxies. The Kolmogorov-Smirnov (KS) test shows that the possibility of the two distributions being the same is only 0.024. Therefore, the dust attenuation in paired galaxies is not the same as in isolated galaxies. The effect of this difference in computing SSFRs is discussed in Section \ref{sec_ssfr}. We now analyze the dependence of dust attenuation on different physical properties for our pair and control samples. 
%%%Dust extinction%%%

\subsection{Dependence of dust attenuation on mass}
\label{subsec_dustmassdep}

\begin{table*}
\caption{Average $A_{\rm FUV}$ for galaxies in pair and control samples in four different mass bins.} \label{tab_extm}
\centering
\begin{tabular}{rcccc}
\hline\hline
Mass bins  & \multicolumn{4}{c}{$A_{\rm FUV}$ (mag)} \\
           &   pair  & control   & S in S-S      & S in S-E\\

\hline
9.7 $<\ \log (M/M_{\odot})\ <$ 10.2  &  0.97$\pm$0.12 & 1.18$\pm$0.20 & 0.97$\pm$0.12 &  ...\\
10.2 $<\ \log (M/M_{\odot})\ <$ 10.8 &  2.48$\pm$0.53 & 2.52$\pm$0.69 & 1.91$\pm$0.72 & 3.35$\pm$0.17\\
10.8 $<\ \log (M/M_{\odot})\ <$ 11.2 &  3.37$\pm$0.28 & 2.48$\pm$0.30 & 3.61$\pm$0.34 & 2.76$\pm$0.38\\
11.2 $<\ \log (M/M_{\odot})\ <$ 11.6 &  2.96$\pm$0.61 & 2.02$\pm$0.42 & 3.51$\pm$0.71 & 1.58$\pm$0.23\\
\hline
\end{tabular}
\end{table*}

\begin{figure}
\centering
\resizebox{\hsize}{!}{\includegraphics{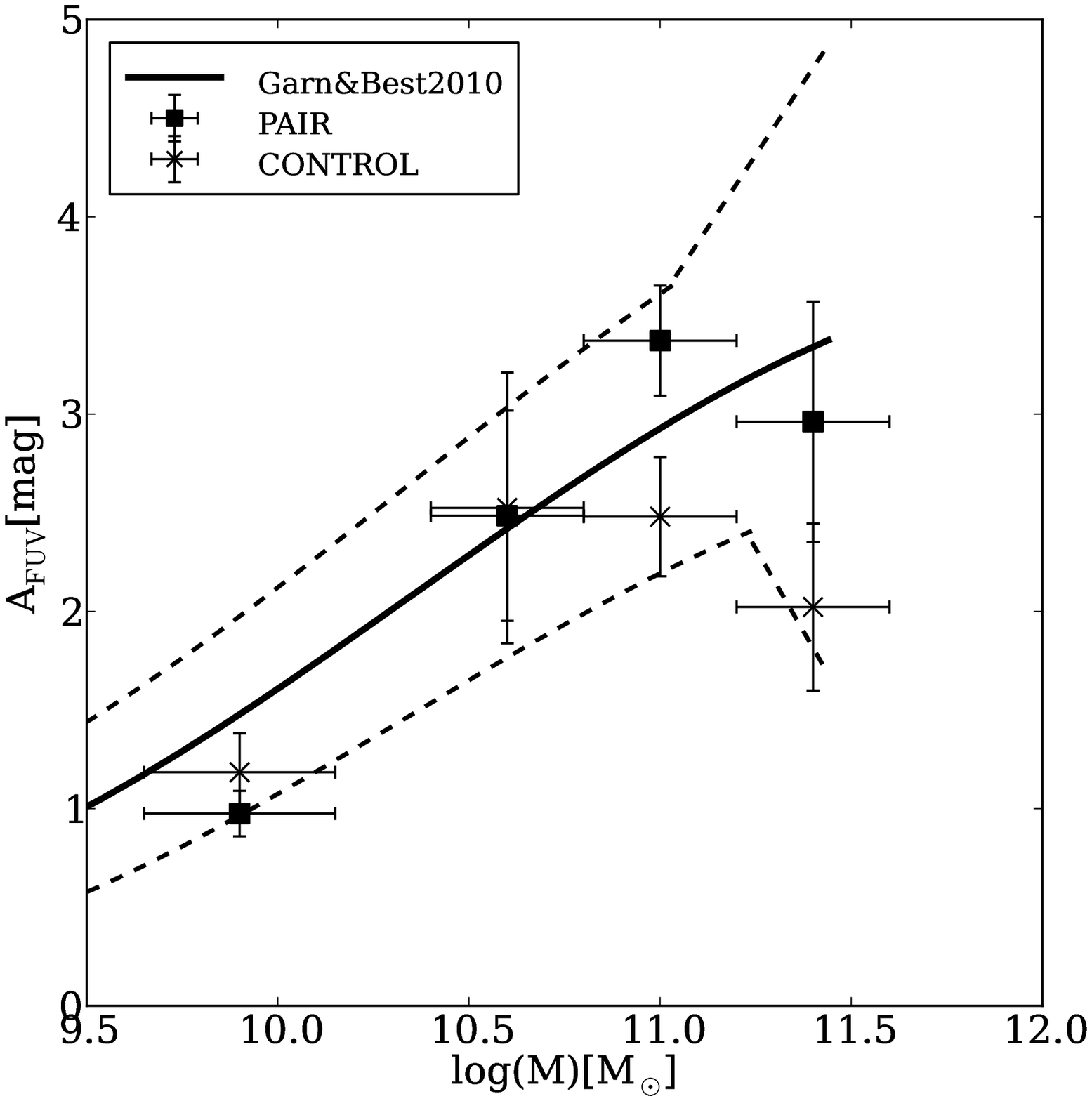}}
\caption{Dependence of $A_{\rm FUV}$ on galaxy stellar mass for spirals in pairs (squares) and for their control galaxies (crosses). The solid line indicates the result given by \cite{garn_best_2010} (modified from $H_{0}=70\ {\rm km\,s^{-1}\,Mpc^{-1}}$ to $H_{0}=75\ {\rm km\,s^{-1}\,Mpc^{-1}}$), and the dashed lines show the 1$\sigma$ uncertainty. \label{fig_extmass}}
\end{figure}

 First we examine the dependence of $A_{\rm FUV}$ on the stellar mass of galaxies. The stellar mass is calculated using the 2MASS {\it K}$_s$-band luminosities (Xu10). We divide the samples into four mass bins using the same bins as Xu10 (see also Section \ref{sec_ssfr}). The average $A_{\rm FUV}$ in each mass bin is given in Table \ref{tab_extm} and shown in Figure \ref{fig_extmass}. The general trend for dust attenuation is that it increases as the stellar mass of a galaxy increases until the stellar mass reaches $\sim 10^{11}M_{\odot}$ for paired galaxies and $\sim 10^{10.5}M_{\odot}$ for isolated galaxies. This correlation between dust attenuation and mass may be due to the correlation between metallicity and dust content \citep{brinchmann_2004}. However, in the more massive bins, dust attenuation seems to decrease.

We compared this trend with previous studies. \cite{kauffmann_stellar_2003} investigated a sample selected with the SDSS \textsl{z}$^{\prime}$ band and found that dust attenuation reaches a maximum at a stellar mass of $10^{10.5}M_{\odot}$ and then decreases as the stellar mass increases. They conclude that when the stellar mass is higher than $10^{10.5}M_{\odot}$, the number of galaxies with old stellar populations rapidly increases, implying that less massive galaxies contain more gas and young stars. A similar trend is found for our \textsl{K}-band selected control galaxies. The mass turn-off of paired galaxies appears higher than in control galaxies, implying that merging processes increase the dust attenuation in galaxies. However, the trend can be due to the selection effect as well. \cite{iglesias-paramo_star_2006} find that dust attenuation is higher for higher stellar mass galaxies in UV-selected samples, whereas it is lower for higher stellar mass galaxies in IR-selected samples. The trend we find for control galaxies can be explained by the sample being \textsl{K}-band selected and therefore containing both kinds of galaxies. 

\cite{garn_best_2010} have investigated a sample of nearby star-forming galaxies and given the dependence of dust attenuation of H$\alpha$ luminosity $A_{\textrm{H}\alpha}$ on stellar mass. Using $A_{\rm FUV}/A_{\textrm{H}\alpha}=1.68$ \citep{meurer_evidence_2009}, we overplot their result in Figure \ref{fig_extmass} to give a comparison of our result with this more global result derived from a larger sample. Our result is consistent with theirs to within 1$\sigma$ uncertainty.
 
In more massive bins, the difference in dust attenuation between the paired and control galaxies becomes larger, implying that more massive galaxies are affected more strongly by merger process.

%Dust mass dependence
\subsection{Dust attenuation in S-S and S-E pairs}
\label{subsec_dustssse}

In the pair sample, there are 26 spirals in S-S pairs, and nine spirals in S-E pairs. The mean $A_{\rm FUV}$ of spirals is $2.89\pm0.31$ mag in S-S pairs and $2.63\pm0.30$ mag in S-E pairs. The KS test shows that for spirals in S-S pairs the probability that the distribution of dust attenuation is the same as for control galaxies is 0.031, while for spirals in S-E pairs the probability is 0.603. Therefore, the $A_{\rm FUV}$ enhancement is mainly contributed by spirals in S-S pairs. It should be noted that the number of spirals in S-E pairs is small, and may therefore lack statistical significance. More data are needed to investigate whether the SSFRs are enhanced in spirals in S-E pairs. 

\begin{figure}
\centering
\resizebox{\hsize}{!}{\includegraphics{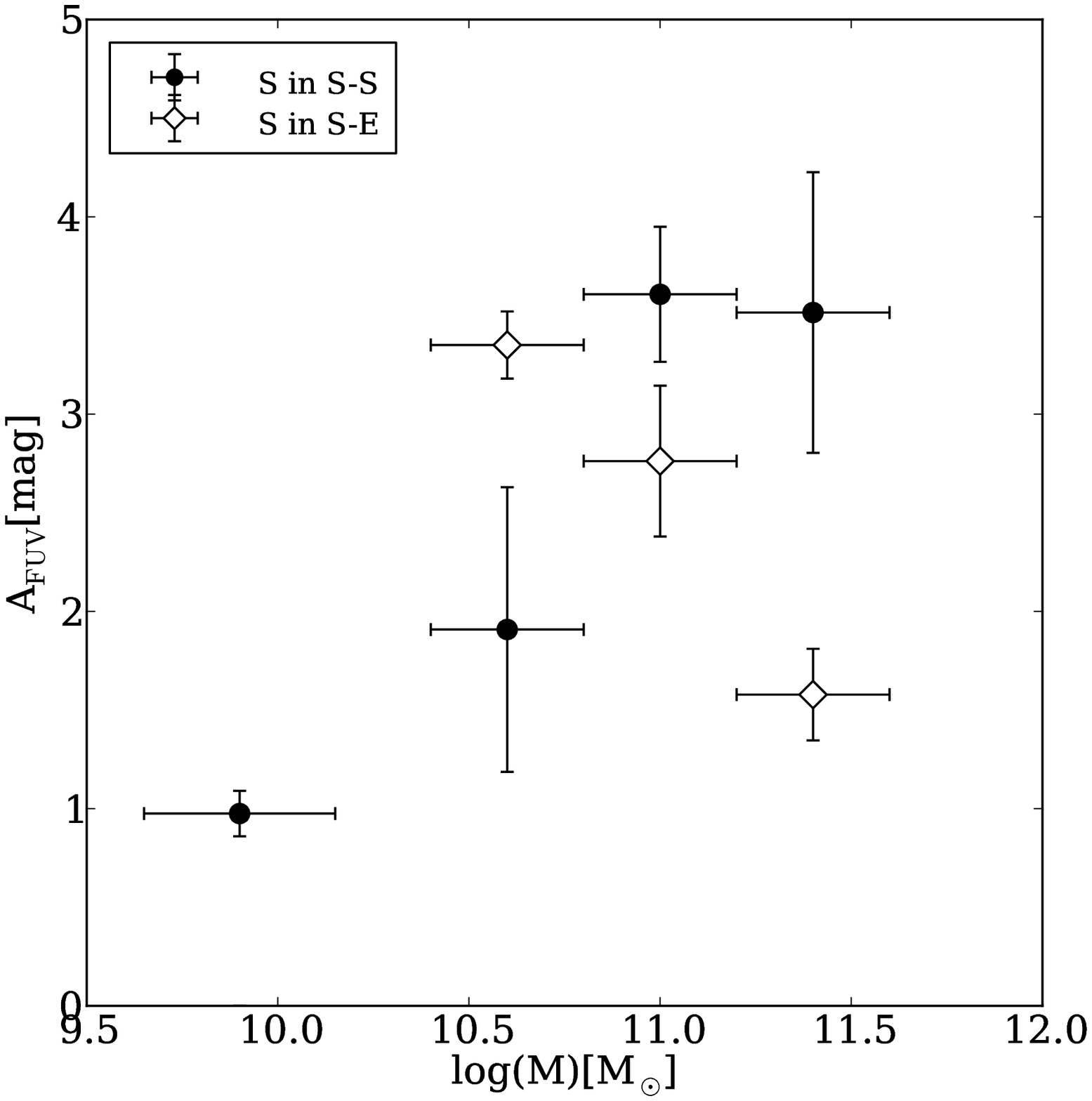}}
\caption{Dependence of $A_{\rm FUV}$ on galaxy stellar mass for non-AGN spirals in S-S pairs (circles) and S-E pairs (diamonds). \label{fig_extmassMor}}
\end{figure}

At high-mass end Figure \ref{fig_extmassMor} shows that the average $A_{\rm FUV}$ of spirals in S-S pairs is apparently larger than S-E pairs. This result may indicate that an S-S interaction extends the duration of gas supply and makes massive galaxies able to have rich gas content, while an S-E interaction dissipates gas and makes massive galaxies more gas poor. The decrease in $A_{\rm FUV}$ in S-E pairs presumably reflects that during an interaction with a hot-gas halo in early-type galaxies, late-type galaxies lose cold gas through hydrodynamic effects such as ram pressure stripping, viscous stripping, and thermal evaporation \citep{park_choi_2009}. 

\begin{figure}
\centering
\resizebox{\hsize}{!}{\includegraphics{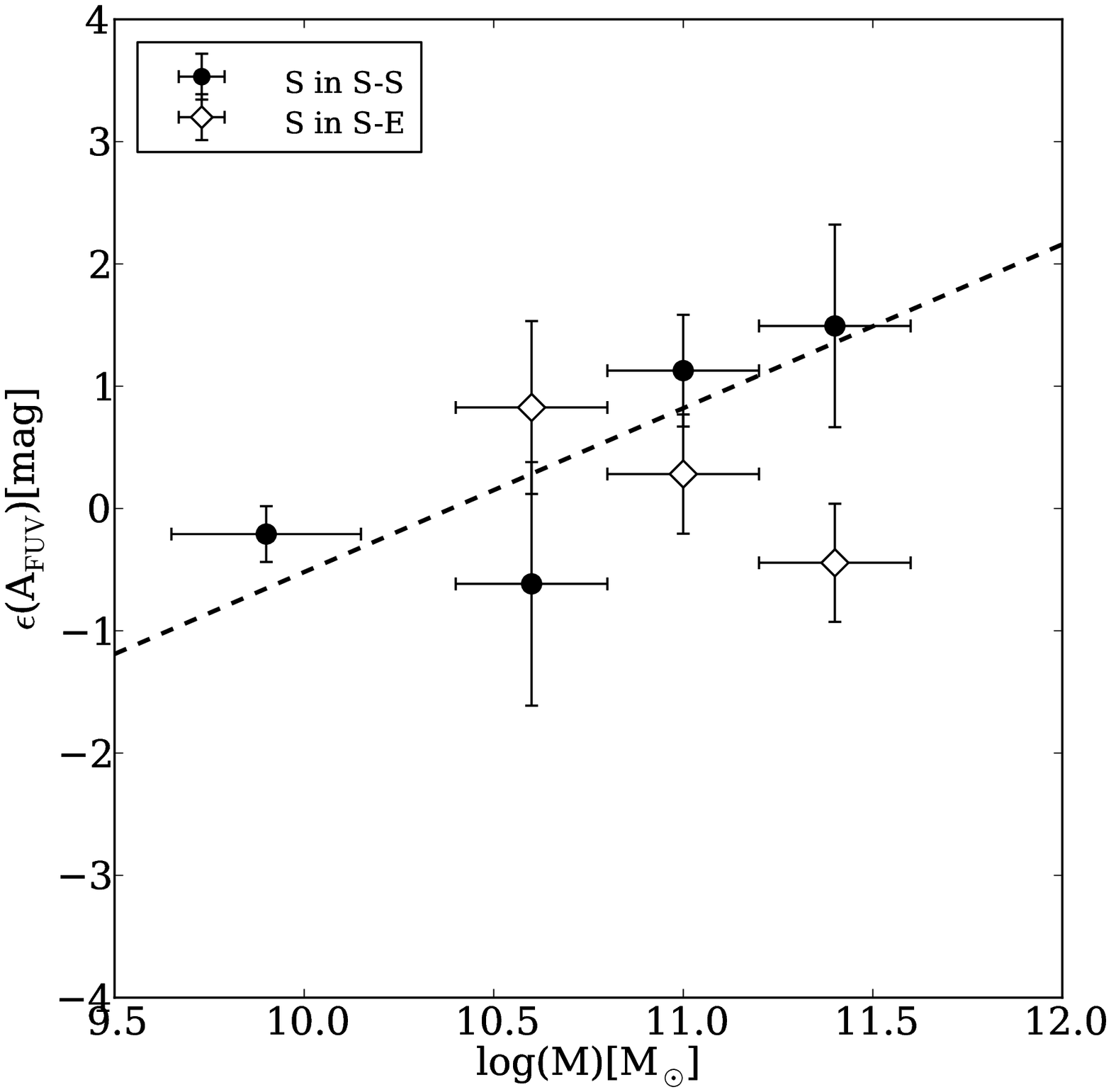}}
\caption{Dependence of $A_{\rm FUV}$ enhancement on galaxy stellar mass for spirals in S-S pairs (dots) and S-E pairs (diamonds). The dashed line is the linear regression for spirals in S-S pairs. \label{fig_extenh}}
\end{figure}

An enhancement parameter $\epsilon(A_{\rm FUV})$ is defined as
\begin{equation}
\epsilon (A_{\rm FUV}) = A_{\rm FUV}^{\rm pair}\textrm{[mag]} - A_{\rm FUV}^{\rm control}\textrm{[mag]}.
\end{equation}
The $\epsilon$ for paired and control galaxies are plotted in Figure \ref{fig_extenh}. The enhancement for spirals in S-S pairs increases as the stellar mass of galaxies becomes higher, while the enhancement of spirals in S-E pairs shows the opposite trend. These trends can be quantitatively described by the linear regressions
\begin{eqnarray}
\label{equ_dustenss}
\nonumber \lefteqn{<{\epsilon}^{S-S}(A_{\rm FUV})\textrm{[mag]}>} \\
 &=& (-13.92\pm 7.33) + (1.34\pm 0.68)\,\log M [M_{\odot}]
\end{eqnarray}
and
\begin{eqnarray}
\label{equ_dustense}
\nonumber \lefteqn{<{\epsilon}^{S-E}(A_{\rm FUV})\textrm{[mag]}>}\\
 &=& (17.68\pm 1.44) - (1.59\pm 0.13)\,\log M [M_{\odot}]. 
\end{eqnarray}

The trend for spirals in S-E pairs still needs further tests because, as mentioned above, there are only nine galaxies in this sample so the result may not be representative. In the rest of this section, we focus on spirals in S-S pairs.

%Dust SS and SE

%Dust enhancement

\subsection{Dust attenuation in primaries and secondaries}
\label{subsec_dustprisec}

\begin{figure}
\centering
\resizebox{\hsize}{!}{\includegraphics{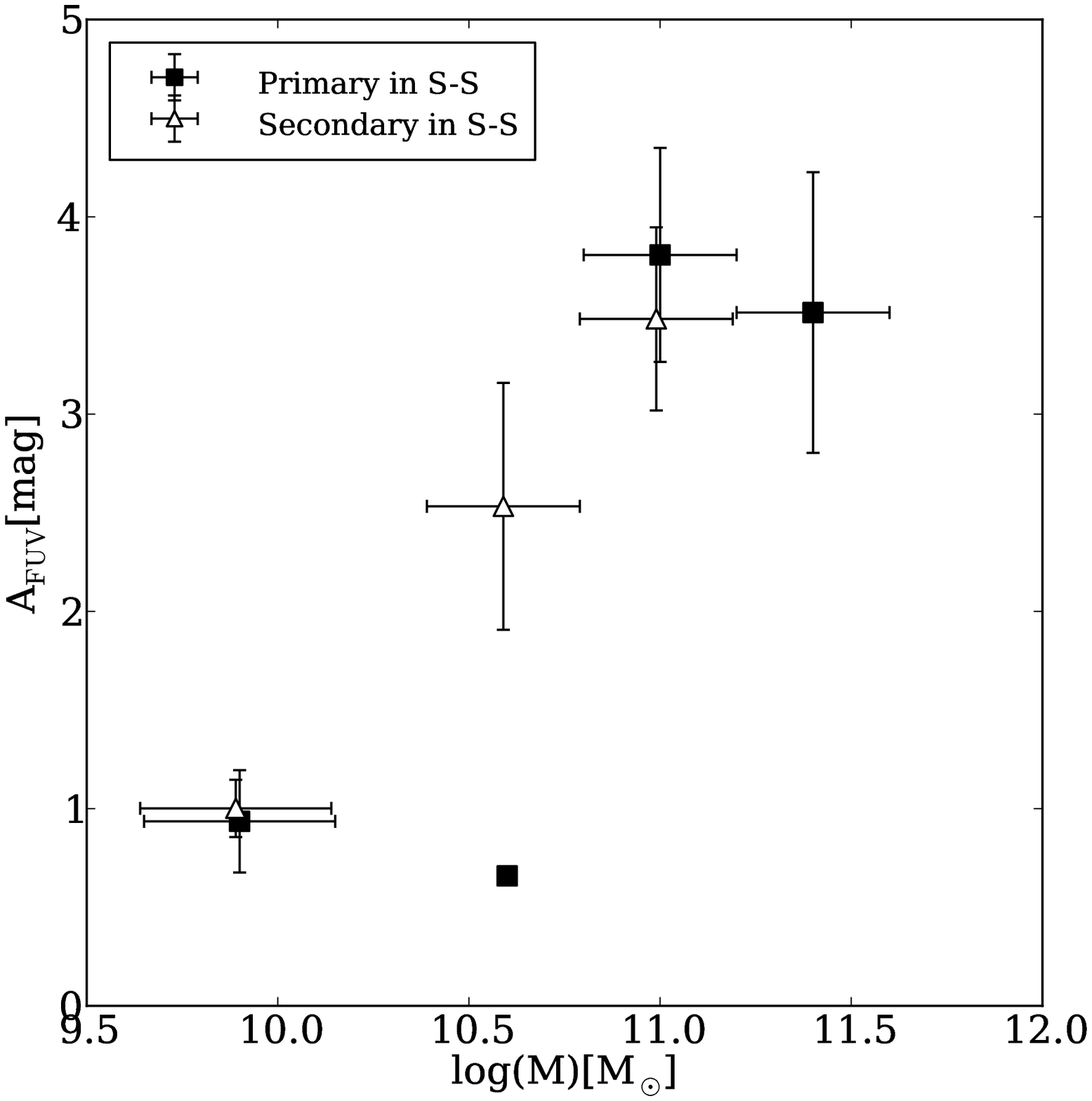}}
\caption{Mean $A_{\rm FUV}$ of primaries (squares) and secondaries (triangles) in S-S pairs in different mass bins. There is only one primary galaxy in the second mass bin. \label{fig_extprisec}}
\end{figure}

In our S-S pairs, the mean $A_{\rm FUV}$ of primaries (13 galaxies) is $3.01\pm0.47$ mag, and that of secondaries (13 galaxies) is $2.76\pm0.41$ mag. The difference is within the standard error. The KS test gives 0.828 as the possibility that these two samples are drawn from the same distribution. Figure \ref{fig_extprisec} shows the mean $A_{\rm FUV}$ of primaries and secondaries in each mass bin. No apparent trend is found.

%Dust pri and sec

\subsection{Dust attenuation and separations}
\label{subsec_dustsep}

\begin{figure}
\centering
\resizebox{\hsize}{!}{\includegraphics{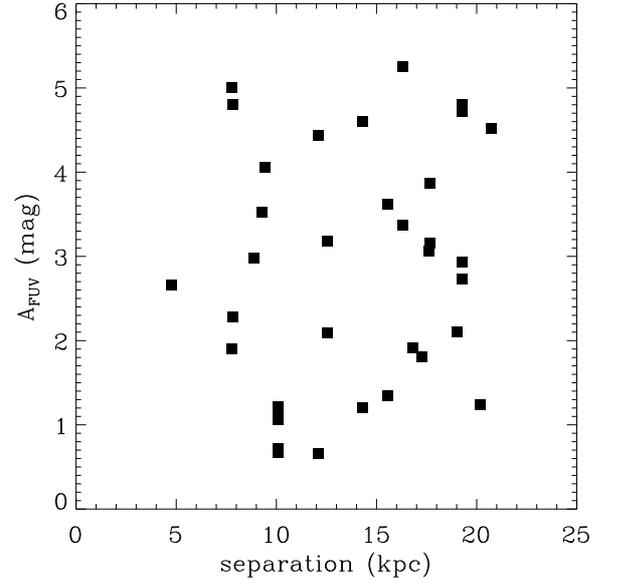}}
\caption{Dependence of $A_{\rm FUV}$ of paired galaxies on the separations.}
\label{fig_sepext}
\end{figure}

A smaller separation between two components of a paired galaxy may indicate a later stage of merging, during which the amount and distribution of dust can be influenced more severely than at an earlier stage. Therefore, the dust attenuation in merger galaxies may depend on the spatial separation between the two components. However, in our sample we find no such dependence (Figure \ref{fig_sepext}). 

We also tested the dependence using the normalized separation parameter SEP defined in Xu10:
\begin{equation}
\label{equ_sep}
  {\rm SEP} = \frac{s}{r_1\ +\ r_2}, 
\end{equation}
where $s$ is the projected separation, and $r_1$ and $r_2$ are the {\it K}-band Kron radii of the primary and secondary from 2MASS, respectively. $s$, $r_1$, and $r_2$ have the same units. For round galaxies, when SEP is smaller than 1 the two components will overlap. The SEP$=$1 is chosen to divide the sample because this value is close to the mean SEP ($1.12\pm0.08$) of the whole sample. It also makes our analysis consistent with Xu10 and allows a convenient comparison between our results and theirs. 

\begin{figure}
\centering
\resizebox{\hsize}{!}{\includegraphics{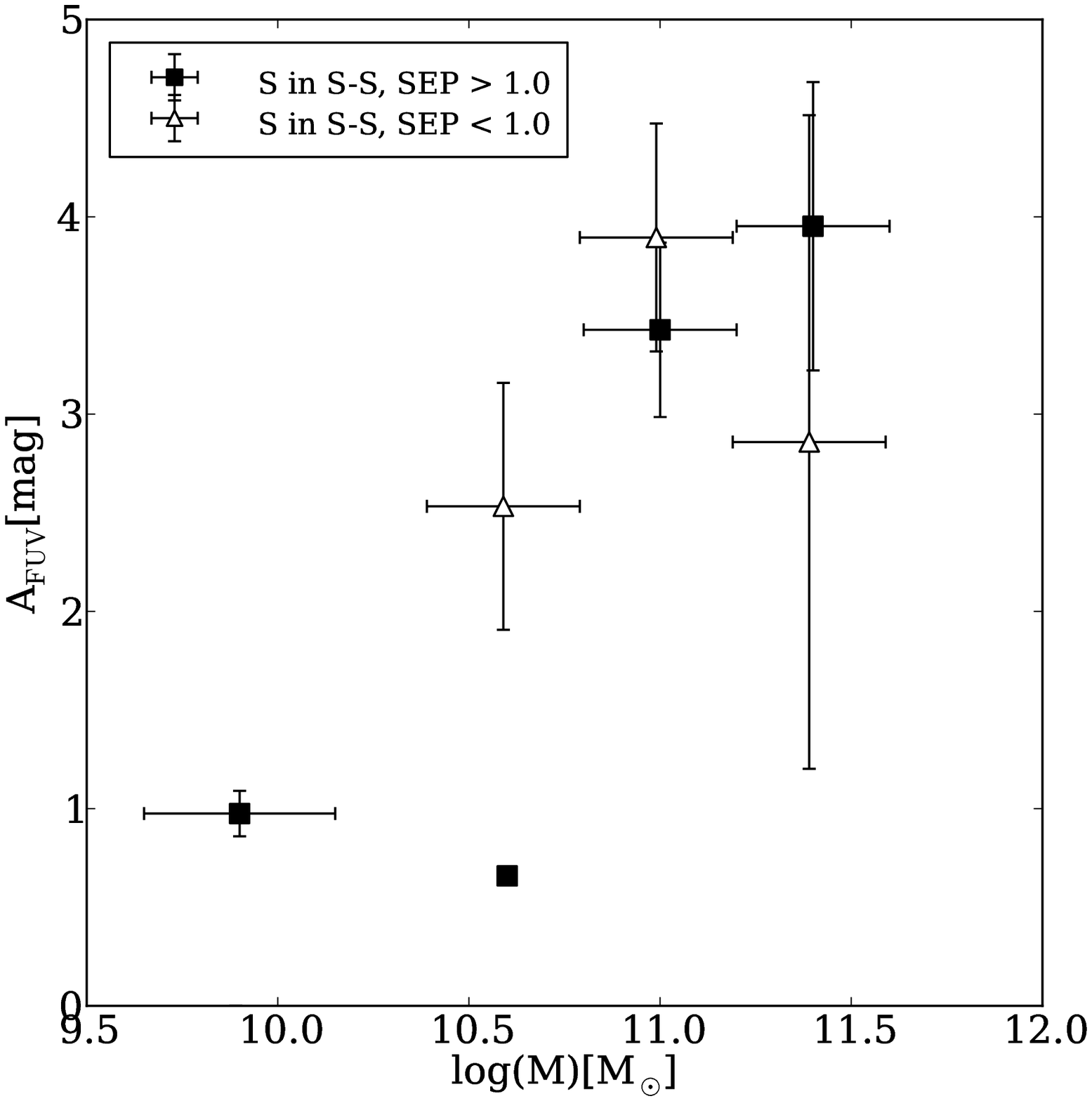}}
\caption{Mean $A_{\rm FUV}$ of non-AGN spirals in S-S pairs with normalized separations (SEP) greater than 1 (squares) and less than 1 (triangles) in different mass bins. There is only one galaxy with SEP$>$1 in the second mass bin. }
\label{fig_extsep}
\end{figure}

The mean $A_{\rm FUV}$ is $2.64\pm0.39$ mag for pairs with SEP greater than 1 (16 galaxies), and $3.36\pm0.48$ mag for pairs with SEP less than 1 (7 galaxies). The difference is within the error. Figure \ref{fig_extsep} plots the mean $A_{\rm FUV}$ for spirals in S-S pairs with SEP greater than 1 and less than 1 in each mass bin. There is an increase in $A_{\rm FUV}$ for galaxies with SEP greater than 1 when the stellar mass of galaxies increases, while hardly any trend can be seen for galaxies with SEP less than 1. It is possible that the gas content can be enriched at intermediate distances, whereas the situation becomes more complicated when the two galaxies come closer (for instance, increasing SFRs may deplete gas), but this result may be biased by the projection effect.

%extinction vs. separation
\subsection{IRX-$\beta$ relation}
\label{subsec_irxb}
\begin{figure}
\centering
\resizebox{\hsize}{!}{\includegraphics{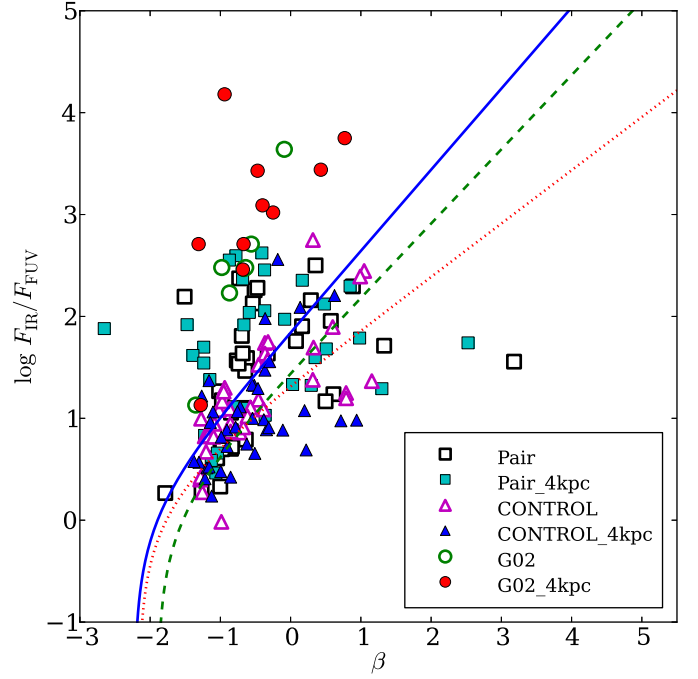}}
\caption{IRX-$\beta$ relation for spirals in the pair sample (squares) and their counterparts in the control sample (triangles). The circles are the results for ULIRGs from \cite{goldader_far-infrared_2002}. Filled symbols represent the quantities within a 4 kpc aperture. The results of \cite{meurer_dust_1999} (solid line), \cite{munoz-mateos_dust_2009} (dotted line), and \cite{takeuchi2012} (dashed line) are plotted as reference.}
\label{fig_irxb}
\end{figure}

IRX is the IR to UV ratio. $\beta$ is the slope of the UV spectrum assuming a power law $F_{\lambda}\varpropto \lambda^{\beta}$, and can be inferred from the difference between the FUV and NUV bands \citep{kong_star_2004}. The IRX-$\beta$ relation has been intensively discussed in many works. Several formulae have been proposed to represent the relation for different types of galaxies. To name a few, \cite{boissier_radial_2007}, \cite{cortese_uv_2006}, \cite{meurer_dust_1999}, \cite{munoz-mateos_dust_2009}, \cite{overzier_dust_2011}, \cite{takeuchi2012}, etc. With FUV, NUV, and IR data, we can examine the IRX-$\beta$ relation in paired galaxies and in their nuclei. In Figure \ref{fig_irxb}, we plot IRX versus $\beta$ for spirals in our pair and control samples, and overplot the relations given by \cite{meurer_dust_1999}, \cite{munoz-mateos_dust_2009}, and \cite{takeuchi2012} as comparisons. Meurer's law is only applicable to starburst galaxies. \cite{takeuchi2012} re-derived the relation for the same sample with newer data and corrected for the aperture effect. \cite{munoz-mateos_dust_2009} investigated the relation for more quiescent galaxies using a SINGS sample.   

Figure \ref{fig_irxb} shows the control galaxies follow these relations closely. In contrast, the locations of the paired galaxies are spread from the region of the quiescent galaxies to the region of \cite{goldader_far-infrared_2002} ultraluminous IR galaxies (ULIRGs). The large scatter for paired galaxies in the IRX-$\beta$ diagram suggests that interaction complicates the physical processes on dust attenuation. Most of the paired galaxies have a similar location to those found by \cite{jonsson_simulations_2006} for luminous mergers using numerical simulations. The positions of the nuclear regions of paired galaxies are similar to those of the ULIRGs in \cite{goldader_far-infrared_2002}. This is consistent with an interaction injecting gas into galaxies, especially in the nuclear regions. The central regions of ULIRGs are more extreme and show the largest deviations from the IRX-$\beta$ laws. These deviations can be explained by these galaxies, especially the central parts, being strongly obscured by dust. Only a very small fraction of UV light can be detected, and thus IRX has very little correlation with $\beta$ \citep{goldader_far-infrared_2002}. 

%irx-beta

\section{SFRs in merger galaxies}
\label{sec_ssfr}

In Section 3, we find that the distributions of dust attenuation in pair and control samples are different and that the attenuation is dependent on morphology and mass. Therefore, the result of Xu10 may be significantly affected by the UV part of SFR that is not included in their analysis. A better way to calculate the SFRs in galaxies is to combine the obscured SFRs and unobscured SFRs \citep[e.g.][]{takeuchi_evolution_2005, buat_dust_2005, cortese_uv_2006}. Assuming a constant burst of star formation for $10^8$ yr and a Salpeter initial mass function \citep[IMF;][]{salpeter_1955}, \cite{buat_local_2007} present SFR calibrations using FUV and IR luminosity as follows:
\begin{eqnarray}
\label{equ_sfruv}
\log {\rm SFR_{FUV}}[M_{\odot}\textrm{yr}^{-1}] = \log L_{\rm FUV}[L_{\odot}] - 9.51\\
\label{equ_sfrir}
  \log {\rm SFR_{IR}}[M_{\odot}\textrm{yr}^{-1}] = \log L_{\rm IR}[L_{\odot}] - 9.75.
\end{eqnarray}

Since the IR emission is not only from dust heated by massive young stars, but also from old stars, the contribution of the latter should be removed. The obscured SFRs (SFR$_{\rm dust}$) are then given by
\begin{equation}
\label{equ_sfrdust}
 {\rm SFR_{dust}} = (1-\eta){\rm SFR_{IR}}.
\end{equation} 

Therefore, the total SFRs are given by
\begin{equation}
\label{equ_totsfr}
  {\rm SFR_{TOT}} = {\rm SFR_{FUV}} + (1-\eta) {\rm SFR_{IR}},
\end{equation}  
where $\eta$ is the fraction of IR emission by old stars. We use $\eta=0.3$ as suggested by \cite{buat_star_1996}. \cite{hirashita_star_2003} give a higher value of 0.4. \cite{bell_estimating_2003} estimated $\eta$ to be $0.32\pm0.16$ for galaxies with $L_{\rm IR}<10^{11}L_{\odot}$, and $0.09\pm0.05$ for galaxies with $L_{\rm IR}>10^{11}L_{\odot}$. \cite{buat_spectral_2011} find an average value of $0.17\pm0.10$ for a sample of star-forming galaxies. It should be stressed that the difference caused by using a different value of $\eta$ is not significant. Although \cite{bell_estimating_2003} gives an apparently lower $\eta$ for galaxies with $L_{\rm IR}>10^{11}L_{\odot}$, only three galaxies with such high luminosities are included in our sample. Even if $\eta=0.09$ is taken, the total SFRs are at most 0.1 dex greater than the case of $\eta=0.3$, which is within the uncertainty.

\begin{figure}
\includegraphics[width=\linewidth]{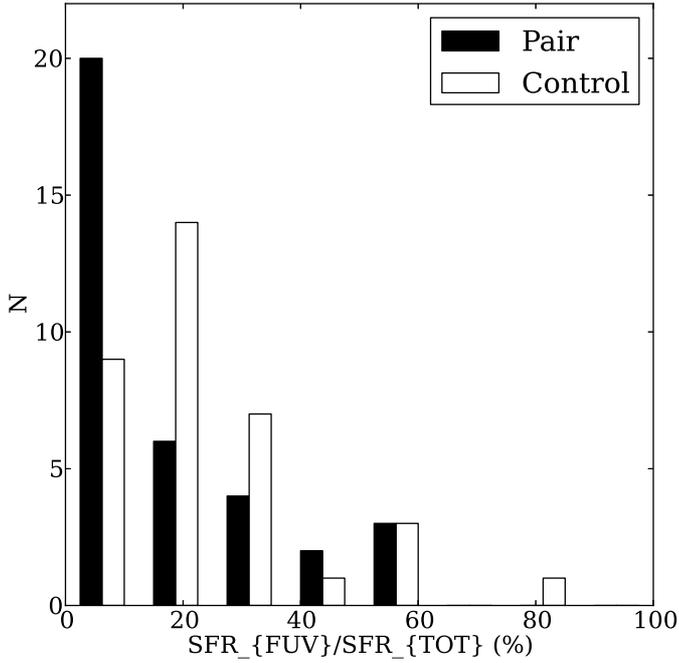}
\caption{Histograms of SFR$_{\rm FUV}$/SFR$_{\rm TOT}$ distributions for spirals in the pair and control samples. \label{fig_frac}}
\end{figure}

For the 39 spirals in pairs and their control galaxies, SFR$_{\rm TOT}$, SFR$_{\rm FUV}$, SFR$_{\rm dust}$, and the fraction of total SFRs contributed by FUV parts are listed in Table \ref{tb3}. The distributions of the fractions of SFR$_{\rm FUV}$ for paired and control galaxies are quite different (Figure \ref{fig_frac}; the KS test gives a 0.024 probability that these two distributions are the same). The SFR$_{\rm FUV}$ contributes from several percent to as high as 80 percent, which can be sufficient to affect the distribution of the total SFRs. 

\begin{figure}
\includegraphics[width=\linewidth]{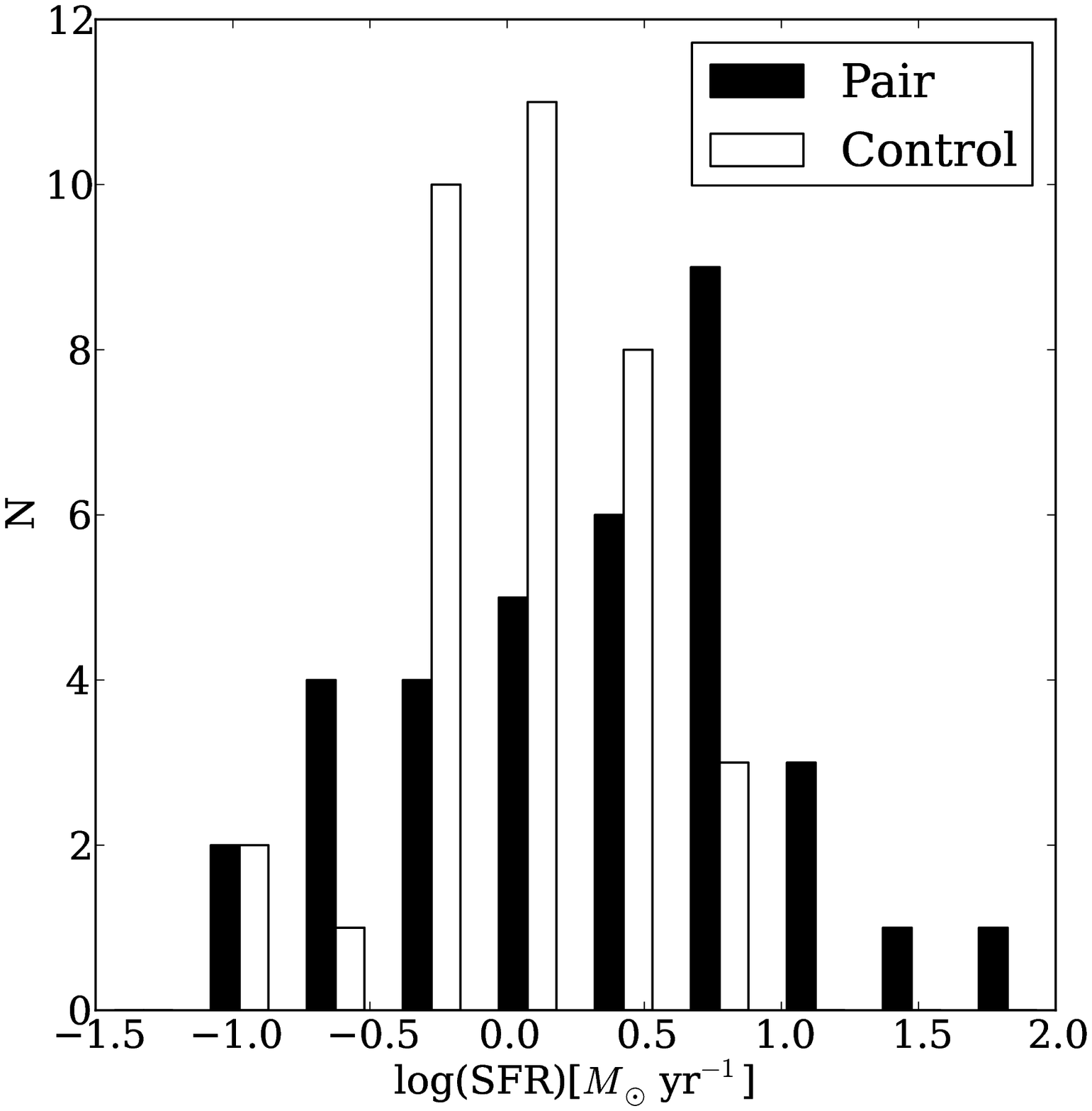}
\caption{Histograms of SFR distributions for spirals in the pair and control samples. \label{sfrtotHist}}
\end{figure}

Figure \ref{sfrtotHist} shows a comparison of histograms of SFRs for the pair and control samples. The histograms show a significant excess of paired galaxies in the high SFR end. There are five galaxies with $\log$(SFR) larger than one in the pair sample, but none in the control sample. The KS test gives 0.049, quite a low value for the null hypothesis that these two samples are drawn from the same population. The mean $\log$(SFR) in pair and control samples are $0.32\pm0.12$ and $0.06\pm0.07$, respectively. In paired galaxies, the contribution from old stars should be less than 0.3 because of the higher star formation activity. Therefore, the SFR in paired galaxies is slightly more than we estimated. That both dust attenuation and SFRs are found enhanced in paired galaxies is consistent with the correlation between SFRs and $A_{\rm FUV}$ found by previous studies \citep[e.g.][]{brinchmann_2004,iglesias-paramo_uv_2004}.

\begin{figure}
\includegraphics[width=\linewidth]{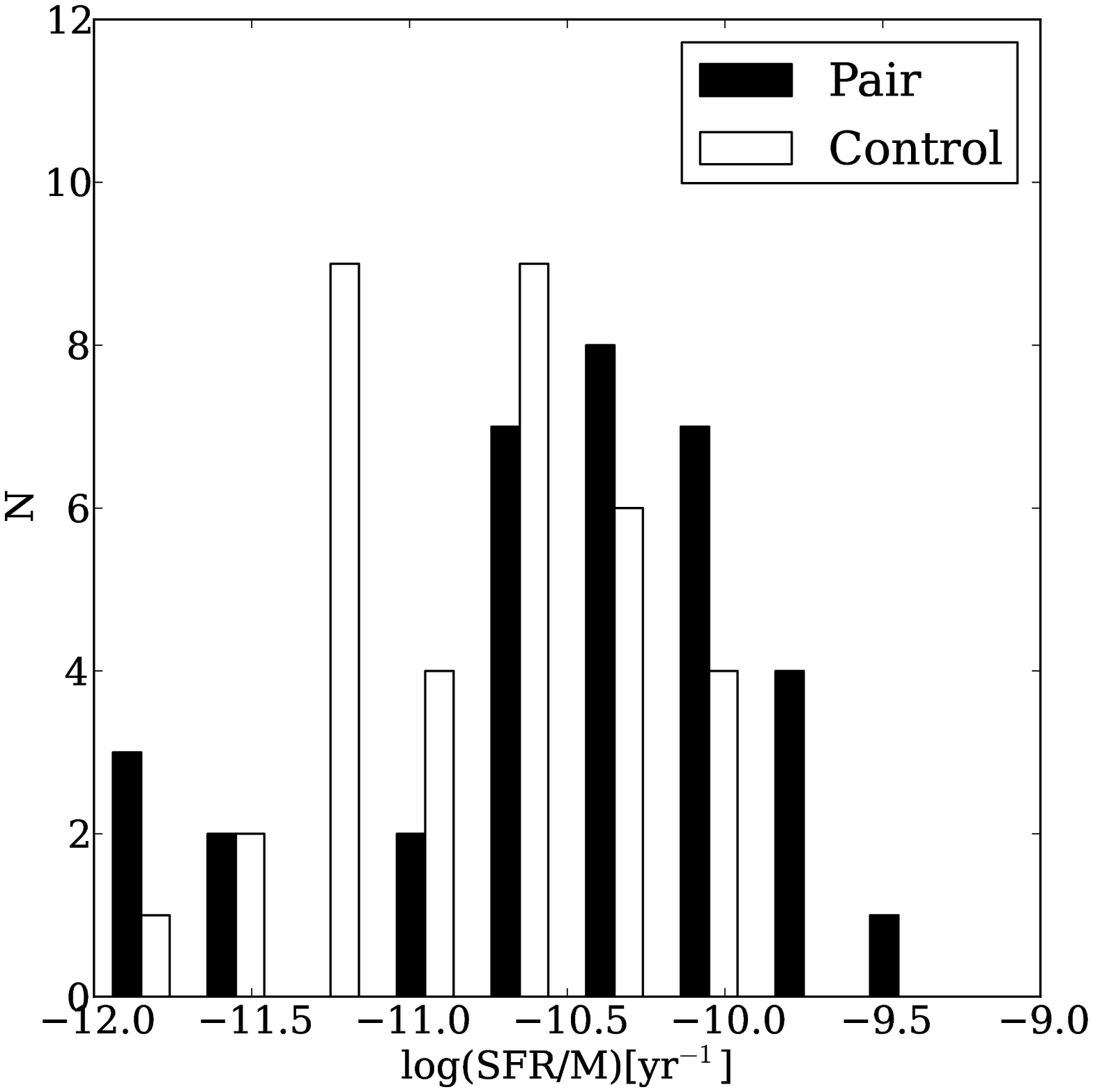}
\caption{Histograms of SSFR distributions for spirals in the pair and control samples. \label{ssfrtotHist}}
\end{figure}

SSFR is the SFR normalized by the stellar mass $M$:
\begin{equation}
\label{equ_totssfr}
  {\rm SSFR}[\textrm{yr}^{-1}] = \frac{{\rm SFR}[M_{\odot}\textrm{yr}^{-1}]}{M[M_{\odot}]}.
\end{equation}
Table \ref{tb3} lists the results of SSFR$_{\rm TOT}$ (${\rm SFR}_{\rm TOT}/M$) for paired and control galaxies. The total SSFR distributions of pair and control samples are shown in Figure \ref{ssfrtotHist}. (Hereafter, we abbreviate SSFR$_{\rm TOT}$ as SSFR.) The two distributions show significant differences from each other. The KS test yields a probability of 0.024 that the two distributions are the same. The mean $\log$SSFR is $-10.54\pm0.11$ for the pair sample, and $-10.79\pm0.08$ for the isolated control sample. Therefore, the paired galaxies show an apparent enhancement of SFRs and SSFRs. This result is consistent with Xu10's conclusion. 

%%add

\subsection{Dependence of SSFR enhancement on mass}
\label{subsec_massdep}

\begin{table*}
\caption{Average SSFRs for pair and control galaxies in four different mass bins.} \label{tab_ssfrm}
\centering
\begin{tabular}{rcccc}
\hline\hline
Mass bins  & \multicolumn{4}{c}{$\log$SSFR (yr$^{-1}$)} \\
           &  pair   & control   & S in S-S      & S in S-E\\
\hline
9.7 $<\ \log (M/M_{\odot})\ <$ 10.2  & -10.28$\pm$0.15 & -10.23$\pm$0.08 & -10.28$\pm$0.15 & ... \\
10.2 $<\ \log (M/M_{\odot})\ <$ 10.8 & -10.36$\pm$0.05 & -10.78$\pm$0.21 & -10.38$\pm$0.08 & -10.33$\pm$0.03\\
10.8 $<\ \log (M/M_{\odot})\ <$ 11.2 & -10.62$\pm$0.20 & -10.86$\pm$0.10 & -10.38$\pm$0.21 & -11.24$\pm$0.34\\
11.2 $<\ \log (M/M_{\odot})\ <$ 11.6 & -10.67$\pm$0.21 & -11.03$\pm$0.18 & -10.65$\pm$0.30 & -10.72$\pm$0.16\\
\hline
\end{tabular}
\end{table*}

\begin{figure*}
\includegraphics[width=\linewidth]{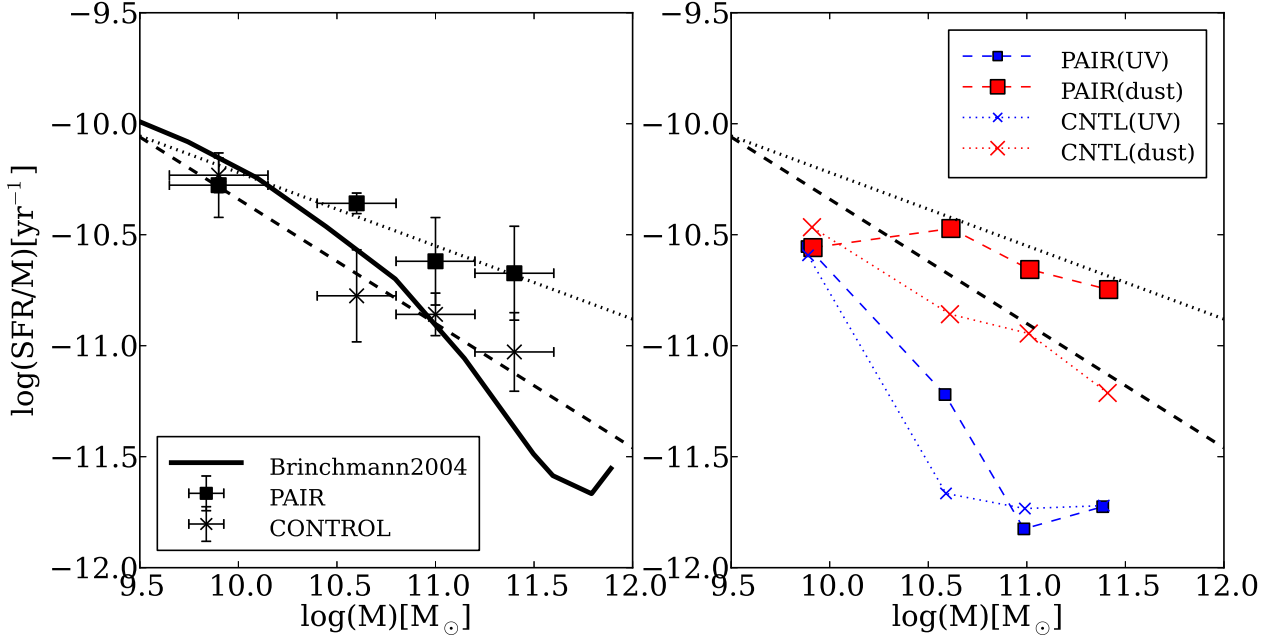}
\caption{Left: Dependence of SSFRs on galaxy stellar mass for paired galaxies (squares) and for the control sample (crosses). The solid line is the result given by \cite{brinchmann_2004}, modified for initial mass function (IMF) and $H_{0}$. Right: Similar to left, but divided into UV (small symbols) and dust (large symbols) parts. The dotted line shows the linear regression of the SSFRs-mass relation for spirals in paired galaxies, and the dashed line for control galaxies.}
\label{massDep}
\end{figure*}

The mass dependence of SSFR enhancement is examined following Xu10's method. The galaxies are divided into four groups with different masses (Table \ref{tab_ssfrm}). The average SSFR versus mass of each group is plotted in Figure \ref{massDep}. The result of \cite{brinchmann_2004} is overplotted to give a comparison. Our result shows good consistency with their result, which was derived from a more generally selected sample.

Figure \ref{massDep} shows that the more massive galaxies have stronger enhancement of SSFR, implying that massive galaxies affect each other more severely. Xu10 find that the SSFRs of spirals in pairs are nearly constant; however, here a decreasing trend is also found for paired galaxies although not as apparent as for the control sample:
\begin{eqnarray}
\label{equ_pssfrm}
\nonumber \lefteqn{{\rm SSFR^{pair}}[\textrm{yr}^{-1}]}\\
 &=& (-7.28\pm 0.75) - (0.30\pm 0.07) \log M[M_{\odot}], 
\end{eqnarray}
and
\begin{eqnarray}
\label{equ_cssfrm}
\nonumber \lefteqn{{\rm SSFR^{ctrl}}[\textrm{yr}^{-1}]}\\
 &=& (-4.94\pm 1.08) - (0.53\pm 0.10) \log M[M_{\odot}].
\end{eqnarray} 
The decreasing trends for both control and paired galaxies are shown in Figure \ref{massDep}. The difference between this result and Xu10's is due to the FUV contribution. The righthand panel of Figure \ref{massDep} presents the dust-obscured and unobscured parts of SFRs. Although the dust-obscured SSFR (SSFR$_{\rm dust}$) is almost constant, the SSFR indicated by FUV (SSFR$_{\rm FUV}$)  becomes less as the mass increases. 

\begin{figure*}
\includegraphics[width=\linewidth]{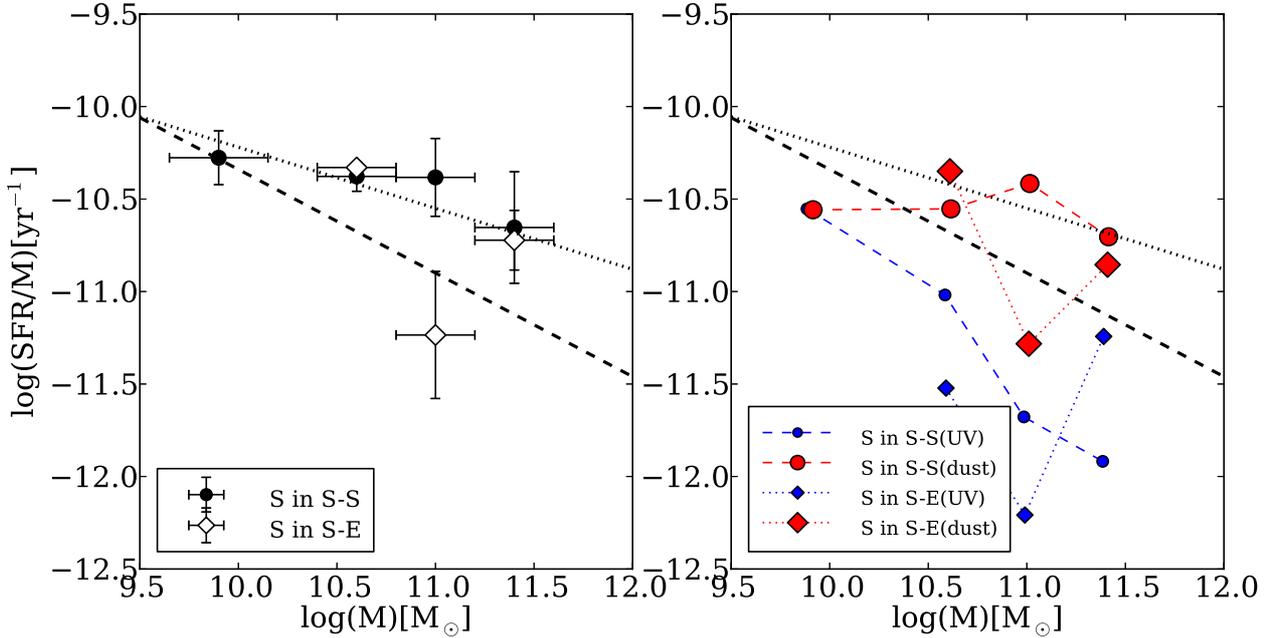}
\caption{Left: Dependence of SSFRs on galaxy stellar mass for spirals in S-S pairs (dots) and S-E pairs (diamonds). Right: Similar to left, but divided into UV (small symbols) and dust parts (large symbols). The lines are the same as in Figure \ref{massDep}.}
\label{massDepMor}
\end{figure*}

\subsection{SSFR Enhancement in S-S and S-E pairs}
\label{subsec_ssse}

\begin{figure*}
\includegraphics[width=\linewidth]{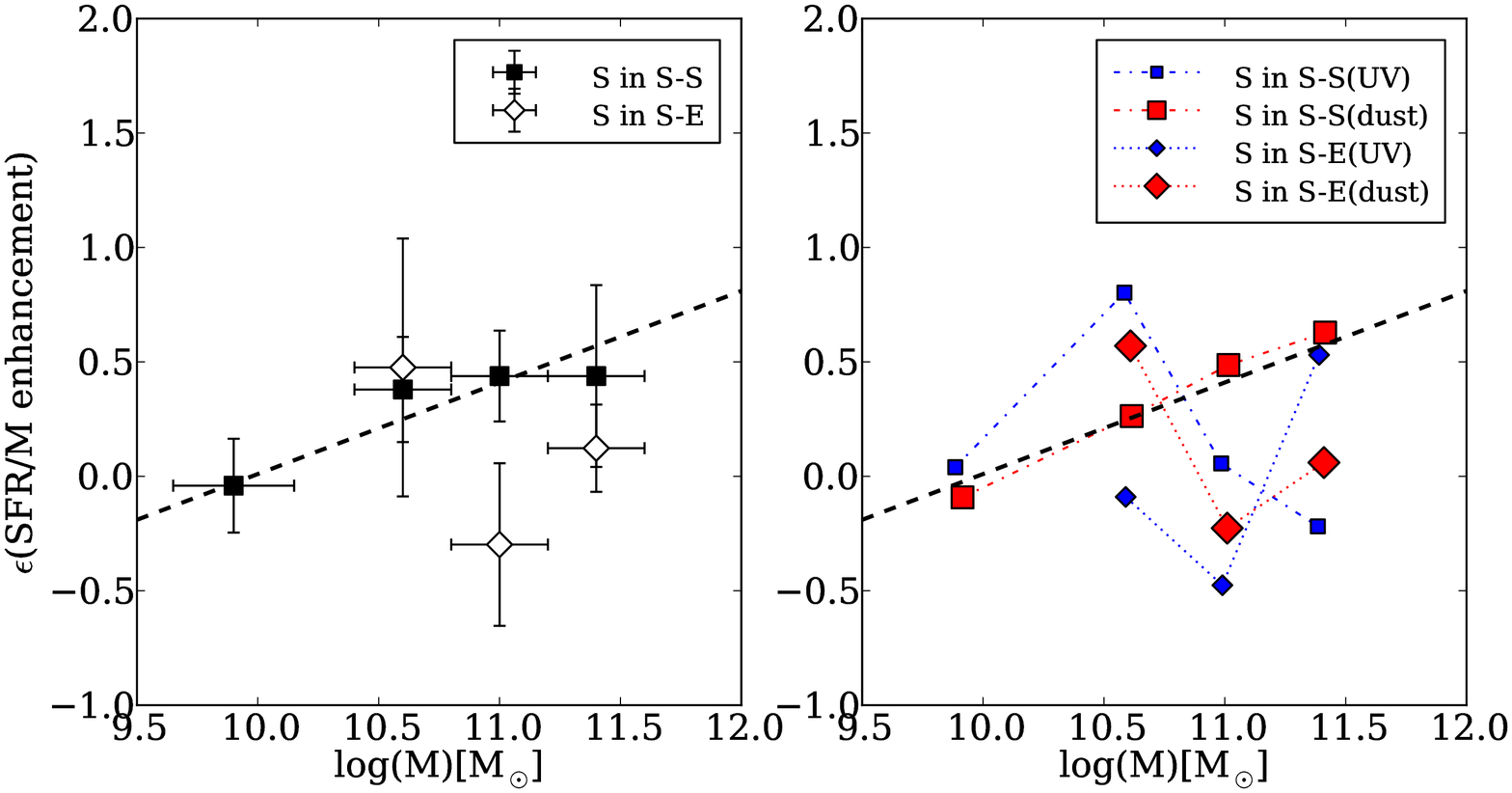}
\caption{Left: Dependence of SSFR enhancement on galaxy stellar mass for spirals in S-S pairs (squares) and S-E pairs (diamonds). Right: Similar to left, but divided into UV (small symbols) and dust parts (large symbols). The dashed line shows the linear regression for spirals in S-S pairs.}
\label{enhance}
\end{figure*}
As mentioned in Section \ref{subsec_dustssse}, among our spirals there are 26 spirals in S-S pairs and nine spirals in S-E pairs. Xu10 find that only the SSFRs of spirals in S-S pairs are enhanced. After adding the unobscured SFRs, we found the mean $\log$SSFR of spirals in S-S pairs is $-10.41\pm0.12$ and that of spirals in S-E pairs is $-10.92\pm0.22$. The KS test shows the possibility of the distributions of SSFRs of spirals in S-S pairs and in their control galaxies are drawn from the same sample is 0.031, and the possibility is 0.957 for spirals in S-E pairs, consistent with Xu10's result. 

Figure \ref{massDepMor} shows the mass dependence of SSFR for spirals in S-S and S-E pairs. Spirals in S-S pairs follow the trend described by Equation (\ref{equ_pssfrm}). On the other hand, the SSFRs of spirals in S-E pairs show no dependence on mass. The righthand panel of Figure \ref{massDepMor} presents the SSFR$_{\rm FUV}$ and SSFR$_{\rm dust}$ for S-S and S-E pairs. For spirals in S-S pairs, the average contribution of SSFR$_{\rm FUV}$ becomes lower at higher mass, implying heavier dust attenuation. For spirals in S-E pairs, neither the SSFR$_{\rm FUV}$ nor SSFR$_{\rm dust}$ has an apparent dependence on mass.  

As defined by Xu10, the SSFR enhancement indicator $\epsilon$ is 
\begin{equation}
\label{equ_ssfren}
  \epsilon = \log {\rm SSFR_{PAIR-S}} - \log {\rm SSFR_{control}}.
\end{equation}
The dependence of $\epsilon$ on mass is plotted in Figure \ref{enhance}. Galaxies in S-S pairs show a clear trend of increasing $\epsilon$ with increasing mass. The correlation between the enhancement in S-S pairs and the stellar mass of galaxies is found to be 
\begin{eqnarray}
\nonumber <\epsilon >_{\rm S+S} &=& (0.01\pm 0.08)\\ 
&&+ (0.40\pm 0.09)\log \frac{M}{(10^{10}M_{\odot})}, 
\end{eqnarray}
which is consistent with Xu10's result within the uncertainties. On the other hand, the $\epsilon$ of spirals in S-E pairs depends little on mass.

The SSFR$_{\rm FUV}$ and SSFR$_{\rm dust}$ for spirals in S-S and S-E pairs are also plotted in Figure \ref{enhance}. The SSFR$_{\rm dust}$ of spirals in S-S pairs has an increasing enhancement as the mass of galaxies increases. However, the SSFR$_{\rm FUV}$ of spirals in S-S pairs does not show this trend. 

\subsection{SSFRs in primaries and secondaries}
\label{subsec_prisec}

\begin{figure*}
\includegraphics[width=\linewidth]{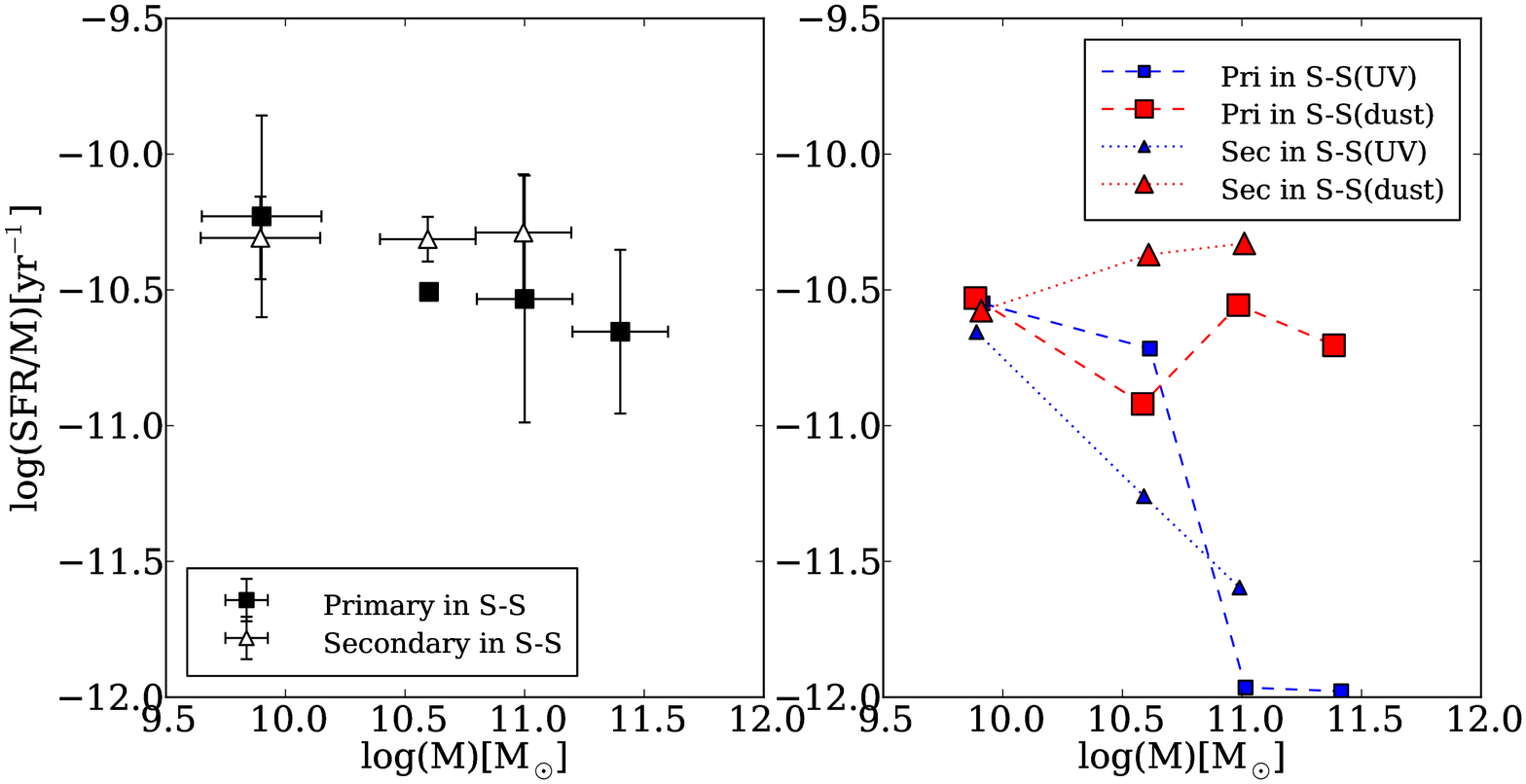}
\caption{Left: Mean SSFRs of primaries (squares) and secondaries (triangles) of S-S pairs in different mass bins. Note that there is only one primary galaxy in the second mass bin. Right: Similar to left, but divided into UV (small symbols) and dust parts (large symbols).}
\label{fig_prisec}
\end{figure*}
\cite{woods_minor_2007} and \cite{ellison_galaxy_2008} have found that secondary components of paired galaxies show stronger enhancements than primary components. Xu10 argue that there is no significant difference between the mean SSFR of primaries and secondaries in any of the mass bins they studied. Here we check the total SSFRs in primaries and secondaries in paired galaxies and in their counterparts in the control sample. The mean $\log$SSFRs of primaries is $-10.53\pm0.20$, and that of secondaries is $-10.30\pm0.13$. Although the secondaries have higher average SSFRs the difference is within the standard error. The KS test for the primaries and secondaries gives a probability of 0.226, so it is hard to conclude whether the distributions are different or not. 

The mass dependence of the SSFRs of primaries and secondaries is also examined (Figure \ref{fig_prisec}). It seems the secondaries have higher SSFRs in more massive bins, but the difference is not significant. No significant difference is found between the primaries and secondaries in each mass bin for SSFR$_{\rm FUV}$ or SSFR$_{\rm dust}$.  

\subsection{Enhancement in one or two components}
\label{holmberg}

KS tests for the secondaries and control galaxies and for the primaries and control sample give probabilities of 0.098 and 0.190, respectively, indicating that both primaries and secondaries in the paired galaxy sample are distributed differently from the control sample. 

\begin{figure}
\includegraphics[width=\linewidth]{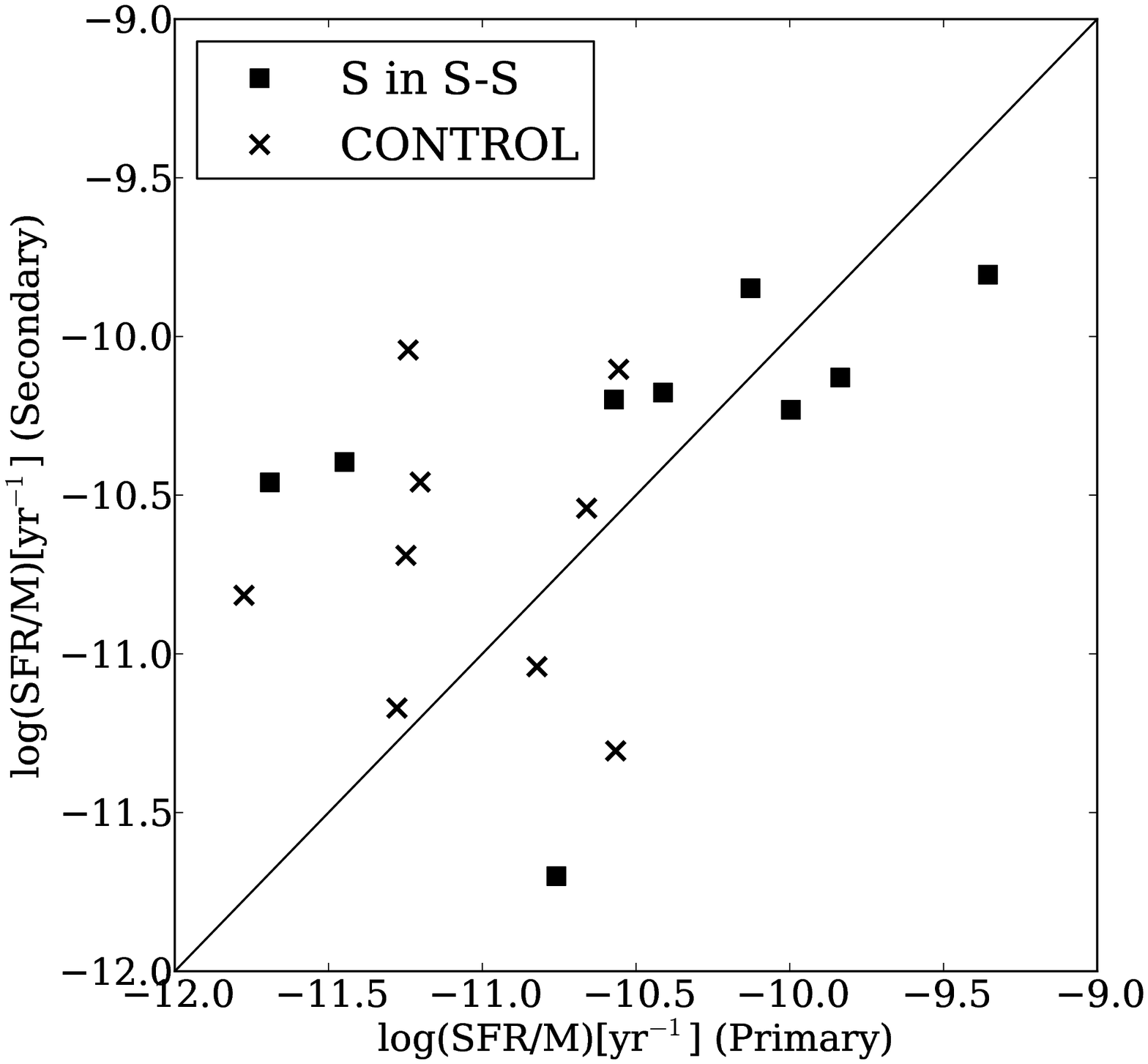}
\caption{Correlation between SSFRs of two components in the S-S pairs with M $\geq\,10^{10.7}$M$_{\odot}$. Squares and crosses represent spirals in the pair and control samples, respectively. }
\label{fig_holm}
\end{figure}

Holmberg found that in galaxy pairs the two components tend to have similar colors, which implies a sign of co-evolution of the two components \citep{holmberg_photographic_1958}. Xu10 examined the Holmberg effect by comparing the SSFRs of the two components in ten massive S-S galaxies. Here we also examine the Holmberg effect in the two components of nine massive galaxy pairs ($\log(M/M_{\odot})>10.7$). Massive pairs are examined because only these pairs show an apparent enhancement of SSFRs, which indicates the merging influence on each component. Figure \ref{fig_holm} shows that there is concordance of the two components. The correlation coefficient $r$ is 0.457 for the two components in galaxy pairs, and is 0.007 for their counterparts in the control sample. Although the correlation between the two components for galaxy pairs is less tight than Xu10's result ($r$ = 0.58), our result still shows a certain level of co-evolution for the two components in pairs. 

\begin{figure}
\includegraphics[width=\linewidth]{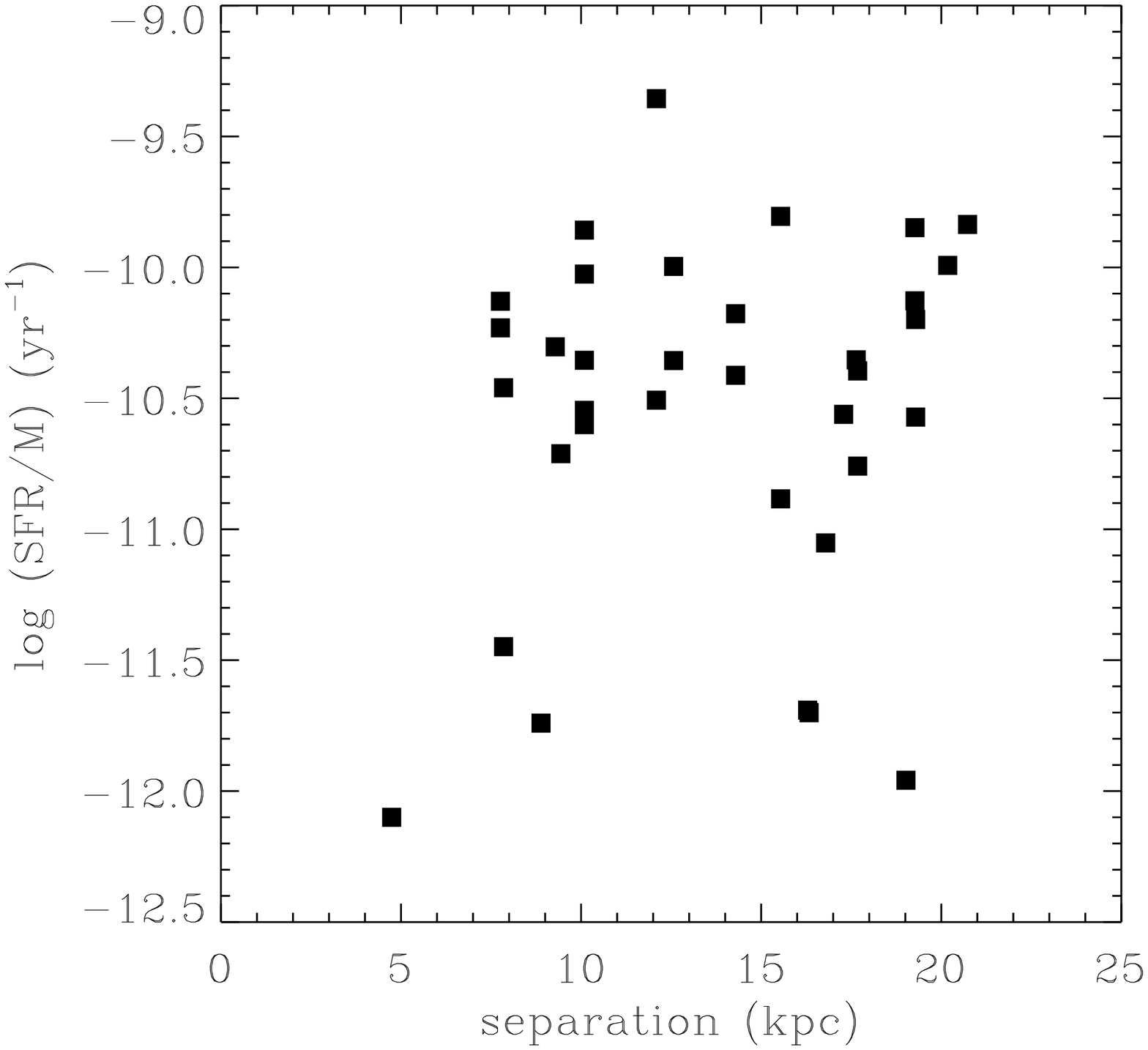}
\caption{Dependence of $\log$ SSFR of paired galaxies on the separations.}
\label{fig_sepssfr}
\end{figure}

\subsection{SSFRs and separations}

\begin{figure*}
\includegraphics[width=\linewidth]{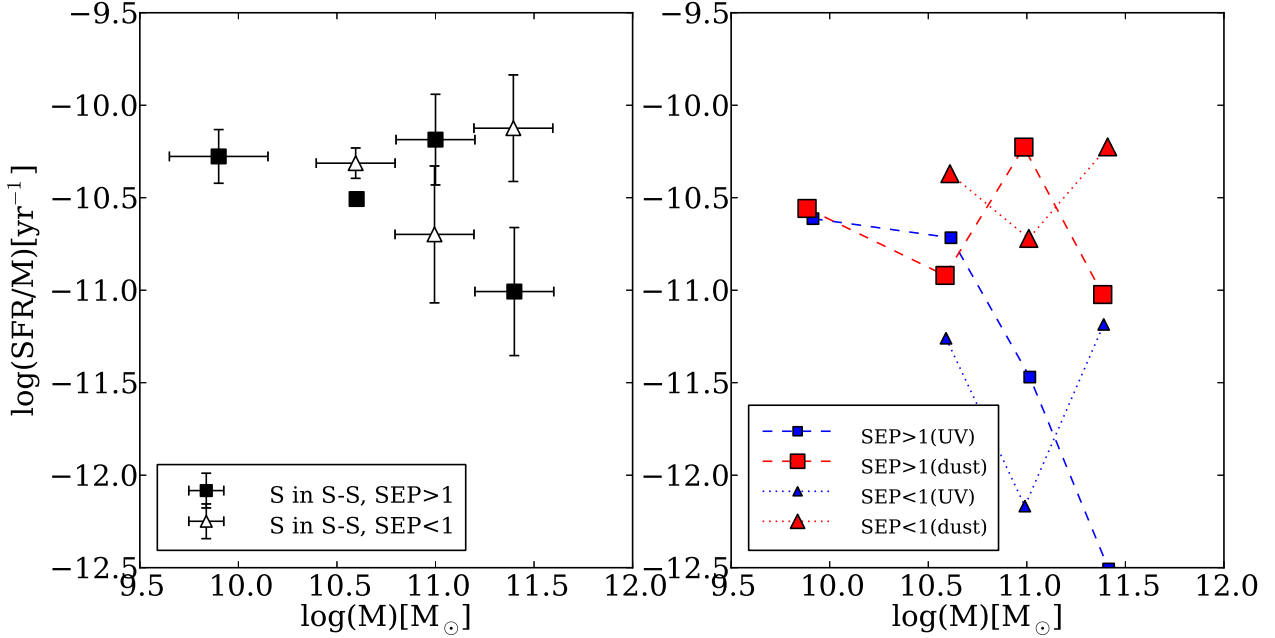}
\caption{Left: Mean SSFRs of spirals in S-S pairs with SEP greater than 1 (squares) and less than 1 (triangles) in different mass bins. There is only one galaxy with SEP$>$1 in the second mass bin. Right: Similar to left, but divided into UV (small symbols) and dust parts (large symbols).}
\label{fig_sep}
\end{figure*}

Figure \ref{fig_sepssfr} shows the scatter plot of SSFRs versus the separations. Hardly any trend can be seen. As in Section \ref{subsec_dustsep}, the average property is also investigated by separating the spirals in S-S pairs into two groups: one with SEP greater than 1, and the other with SEP less than 1. The mean $\log$SSFR for paired galaxies with SEP greater than 1 is $-10.38\pm0.15$, and for those with SEP less than 1 it is $-10.49\pm0.22$. No significant difference is found in the distributions of the two groups (KS test gives 0.802 as the probability that the two distributions are drawn from the same population). Figure \ref{fig_sep} plots the average SSFRs of paired galaxies for these two groups in each mass bin as well as for their obscured and unobscured parts. No systematic trend is found except that the unobscured SSFRs for pairs with SEP larger than 1 seem to decrease as the mass increases. Although previous studies \citep[e.g.][etc.]{xu_solentic_1991, barton_tidally_2000, lambas_2003} conclude that paired galaxies with separation $\lesssim 20h^{-1}$ kpc have stronger enhancement of SFRs than those with separation $\gtrsim 20h^{-1}$ kpc, the separation does not seem to be a determining parameter of star formation activity for mergers with separation $\lesssim 20h^{-1}$  kpc. The outcome can be affected by several conflicting factors. As Xu10 suggest, galaxies with smaller separations may undergo gas depletion due to prolonged star formation activity at the place where the two galaxies overlapped. Also, the projection effect may confuse efforts to probe the actual dependence on the true 3D separations. 

\subsection{SFRs in nuclear regions}
\label{subsec_nuc}

\begin{figure}
\includegraphics[width=\linewidth]{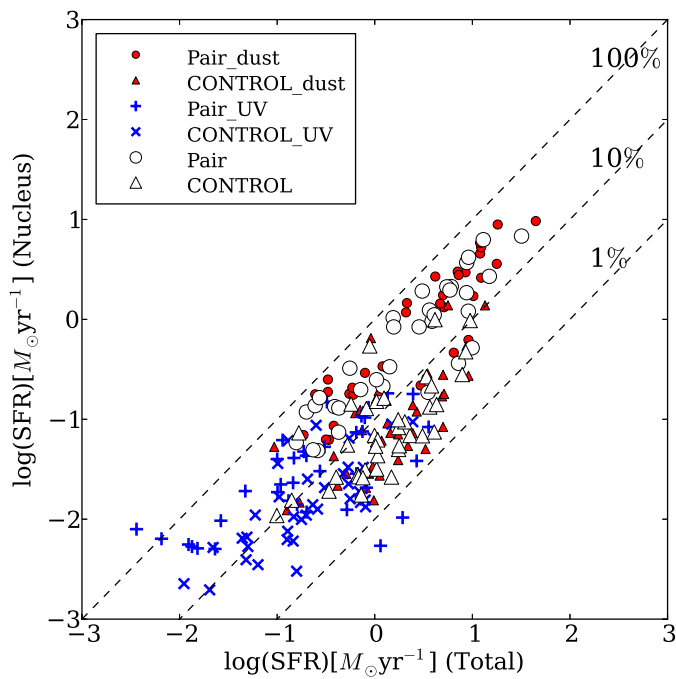}
\caption{Comparison of SFRs included within a 4 kpc aperture and SFRs of a whole galaxy. The lines indicate $1\%$, $10\%$, and $100\%$ nuclear contributions to the total SFRs. Filled circles, pluses, and open circles represent the SFR$_{\rm dust}$, SFR$_{\rm FUV}$, and the total SFRs of paired galaxies, respectively. Filled triangles, crosses, and open triangles are the SFR$_{\rm dust}$, SFR$_{\rm FUV}$, and the total SFRs of control galaxies, respectively.}
\label{fig_nucall}
\end{figure}

From Tables \ref{tb1} and \ref{tb2}, the SFRs inside 4 kpc and 10 kpc are derived. The nuclear contributions to unobscured SFRs (SFR$_{\rm FUV}$), obscured SFRs (SFR$_{\rm dust}$) and their combination are all plotted in Figure \ref{fig_nucall}. The median nuclear contribution of the SFR$_{\rm FUV}$ is $15\%$ for paired galaxies and $6\%$ for control ones. The difference becomes even greater for the SFR$_{\rm dust}$: the average nuclear contribution of SFR$_{\rm dust}$ reaches $33\%$ for paired galaxies, while it becomes $5\%$ for control ones. For the combination of the two parts, the nuclear contribution is $30\%$ for paired galaxies, and is $5\%$ for the control ones.
These results are consistent with the theory that a starburst is triggered in the center of galaxies because of the gas inflow induced by the interaction. Our result is larger compared to the \cite{kennicutt_effects_1987} and \cite{kennicutt_kent_1983} results derived from H$\alpha$ images (in which the central regions contributes $13\%$ and $4\%$). This may be due to the different aperture size: \cite{kennicutt_effects_1987} used an aperture size of $4.7''$, while we used 4 kpc, which corresponds to an average aperture size of $\sim\,9''$ for the pair sample and $\sim\,7''$ for the control sample.

\section{The importance of including the SSFR$_{\rm FUV}$}
\label{sec_importance}

Our results for SSFRs basically agree with the conclusions generated by Xu10, but our findings on dust attenuation in pair and control samples are not similar, as we have shown in Sections \ref{sec_dustatt} and \ref{sec_ssfr}. From Table \ref{tb3}, we can see that in spite of some galaxies having contributions of SSFR$_{\rm FUV}$ over 50\%, the average contributions of SSFR$_{\rm FUV}$ are $17.1$\%$\pm3.1$\% and $22.7$\%$\pm3.1$\% for pair and control samples, respectively. On average, the SSFR$_{\rm FUV}$ contributes much less to the total SSFR than the SSFR$_{\rm dust}$. The average contribution of the SSFR$_{\rm FUV}$ calculated here is less than the mean value given by \cite{takeuchi_evolution_2005} because our sample is near-IR/optical-selected, while their sample is more general and includes objects such as small galaxies with luminous UV emission. Also, their analysis is based on luminosity functions, and therefore volume corrections are accounted for and cannot be compared easily to individual groups. The average SSFR is $-10.54\pm0.11$ in our study, which is very close to what is obtained by Xu10 ($-10.50\pm0.10$). The cosmic star formation density contributed by spirals in paired galaxies is 2.0$\times 10^{-4}$ $M_{\odot}$ yr$^{-1}$ Mpc$^{-3}$ (the spiral fraction is kindly provided by C.K.Xu, private communication), which is consistent with Xu10's result of 2.54$\times 10^{-4}$ $M_{\odot}$ yr$^{-1}$ Mpc$^{-3}$. 

By including SSFR$_{\rm FUV}$, we find similar conclusion to the one in Xu10 that, as the stellar mass of galaxies increases, the enhancement also increases. Our studies of companion morphologies, primaries and secondaries, the Holmberg effect, and the dependence of separations also agree with Xu10's results. Nevertheless, the SSFR$_{\rm FUV}$ in different mass bins is still worth investigating, since the galaxies with high SSFR$_{\rm FUV}$ may fall in the same mass bin and change the total SSFRs distribution severely. Indeed we find the SSFRs for paired galaxies show a slight decrease as the mass of galaxies increases, unlike Xu10's Figure 5 with almost constant SSFRs. One may argue that in more massive bins, the old stellar contribution becomes smaller and the trend may be due to using a constant $\eta$. However, this trend still exists even if we reduce the contribution of the old-stellar population $\eta$ to $9\%$ for the two bins with higher mass. This trend is caused by the decreasing UV contribution as the mass increases, implying that at the low-mass end the UV contribution is quite important and should not be ignored, although in this work the trend is weak and does not affect the calculation of cosmic star formation density. 

\section{Summary and conclusion}
\label{sec_conclusion}

%table 6, average value summary
\begin{table*}\small
\setlength{\tabcolsep}{0.01in}
\caption{Average $A_{\rm FUV}$ and SSFRs for different subsamples of galaxies, where the number in the bracket indicates the number of galaxies in each subsample.} \label{tab_average}
\begin{tabular}{rcccccccc}
\hline\hline
 & Control(35) & pair(35) & S in S-S(26) & S in S-E(9) &\multicolumn{4}{c}{S in S-S} \\
 &             &          &              &             & Primary(13) & Secondary(13) &Sep$>$1(17) & Sep$>$1(9)\\
\hline

$A_{\rm FUV}$ (mag) & $2.20\pm0.21$ & $2.82\pm0.24$ & $2.89\pm0.31$ & $2.63\pm0.30$ & $3.01\pm0.47$ & $2.76\pm0.41$ & $2.64\pm0.39$ & $3.36\pm0.47$\\
$\log$SSFR (yr$^{-1}$)& $-10.79\pm0.08$ & $-10.54\pm0.11$ & $-10.41\pm0.12$ & $-10.92\pm0.22$ & $-10.53\pm0.20$ & $-10.30\pm0.13$ & $-10.38\pm0.15$ & $-10.49\pm0.22$ \\
\hline
\end{tabular}
\end{table*}

%Table 7, the summary of KS test
\begin{table*}
%\setlength{\tabcolsep}{0.02in}
%\tabletypesize{\small}
%\rotate
\centering
\caption{Summary of results of KS tests for $A_{\rm FUV}$ and SSFRs between different subsamples of galaxies.} \label{tab_ks}
\begin{tabular}{rccccc}
\hline\hline
                     &Control vs. Pair&S-S vs. Control &S-E\,\tablefootmark{*} vs. Control &Pri. vs. Sec. &                                             
Sep$>1$ vs. Sep$<1$\,\tablefootmark{*} \\                   
           
\hline

$A_{\rm FUV}$ & 0.024 & 0.031 & 0.603 & 0.828 & 0.579\\
SSFRs         & 0.024 & 0.031 & 0.957 & 0.226 & 0.802\\
\hline
\end{tabular}
\tablefoot{
\tablefoottext{*}{The sample size is small (nine galaxies) so the test results may be biased.}
}
\end{table*}

%nuclear vs. all

We presented FUV and NUV photometry results for a local sample of paired galaxies. By combining the UV and IR data, we investigated the dust attenuation and SFRs in merging spirals and in a control sample of isolated galaxies. Dust attenuation is calculated using UV and IR fluxes and then compared between the pair and control samples. The SFRs indicated by UV are compared with SFRs indicated by IR, and then the UV and IR parts of SFRs are combined to give the total SFRs and SSFRs to study the enhancement of star formation activity in paired galaxies. The results are compared to Xu10's results, which are based on IR images. We also studied the difference in dust attenuation and SSFRs between spiral galaxies in S-S pairs and S-E pairs, between primaries and secondaries, and between paired galaxies with normalized separations SEP greater than 1 and those with SEP less than 1. The KS test result, the mean $A_{\rm FUV}$ and SSFRs in each group are summarized in Tables \ref{tab_average} and \ref{tab_ks}. We come to the following conclusions
\begin{enumerate}
\item Dust attenuation in paired and control galaxies shows different distributions. Paired galaxies tend to have heavier dust attenuation than isolated ones. 

\item The enhancement of dust attenuation depends on the morphology of paired galaxies: spirals in S-S pairs have higher dust attenuation than control galaxies but spirals in S-E pairs do not. The enhancement of dust attenuation for spirals in S-S pairs increases with stellar mass.

\item No systematic difference in dust attenuation is found between primaries and secondaries. Dust attenuation in galaxies with SEP greater than one and those with SEP less than one in S-S pairs do not show significant difference. However, unlike in pairs with SEP less than one, dust attenuation in pairs with SEP greater than one increases with stellar mass.  

\item Our investigation of total SSFRs in paired galaxies confirms Xu10's IR data-only results. The reason is that in our galaxies, the dust obscured SSFRs dominate the total SSFRs at an average level. However, by including SSFRs indicated by UV, we find that SSFRs in paired galaxies also show a decreasing trend as stellar mass increases. 

\item Including the SSFR$_{\rm FUV}$ affects the less massive galaxies most. Nevertheless, this inclusion hardly changes the contribution of paired galaxies to the cosmic SFR in the local universe. 

\item Paired galaxies show a stronger concentration of IR emission and dust attenuation in their nuclear regions than control galaxies do, which is evidence of merger-induced starbursts. 

\item In the IRX-$\beta$ diagram (Figure \ref{fig_irxb}), paired galaxies show larger scatter than control galaxies. The nuclear regions of paired galaxies are located in a similar region to ULIRGs.

\end{enumerate}

Our future work will focus on the spatial symmetry of UV and IR images, as well as on the spectral energy distribution of these galaxies. With the data release of Herschel, submillimeter bands and a larger sample of paired galaxies will be available for further investigation. 

\begin{acknowledgements}
We thank the anonymous referee for helpful suggestions. FTY thanks J. Stone for her valuable comments that improved the clarity of the paper and C. K. Xu for his great help on calculating the cosmic star formation density. FTY, TTT, and YM are partially supported by the Grand-in-Aid for the Global COE Program ``Quest for Fundamental Principles in the Universe: from Particles to the Solar System and the Cosmos'' from the Ministry of Education, Culture, Sports, Science and Technology (MEXT). TTT has been supported by the Grant-in-Aid for the Scientific Research Fund (20740105, 23340046 and 24111707) commissioned by the MEXT of Japan. YM is supported by a Grant-in-Aid for Young Scientists (22684005). VB and DB have been supported by the Centre National des Etudes Spatiales (CNES) and the Programme National Galaxies (PNG). JIP is supported by the grant AYA2007-67965-C03-02, from the Spanish MICINN. This work is based [in part] on observations made with the Spitzer Space Telescope, which is operated by the Jet Propulsion Laboratory, California Institute of Technology under a contract with NASA. This work is also based on observations made with the NASA Galaxy Evolution Explorer. GALEX is operated for NASA by the California Institute of Technology under NASA contract NAS5-98034.
\end{acknowledgements}

\bibliographystyle{aa}
\bibliography{merger_v2}

\begin{appendix}
\section{GALEX GR6 images for pairs in our sample}
\clearpage
\begin{figure*}
  \centering 
  \includegraphics[width=0.7\textwidth]{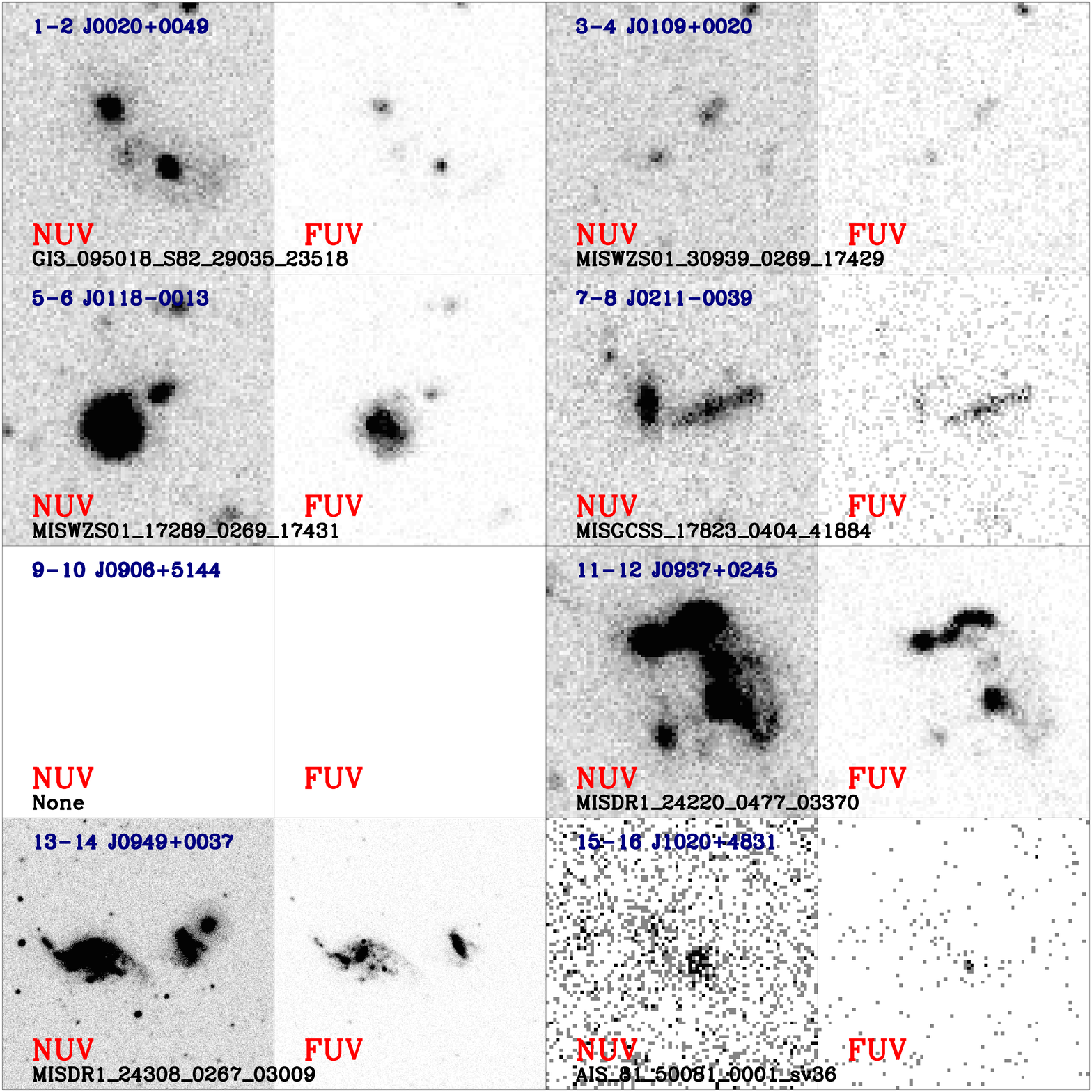}\\
  \caption{GALEX NUV and FUV images of paired galaxies. For pairs 13-14, 37-38, and 51-52, the scale of the images is $7.5'\times 7.5'$. For the other pairs, the scale of the images is $2' \times 2'$. The names of pairs are shown in the upper left of each image, and the names of the GALEX tiles are shown in the lower left. \label{fig1}}
\end{figure*}

\addtocounter{figure}{-1} 
\begin{figure*}
  \centering
  \includegraphics[width=0.7\textwidth]{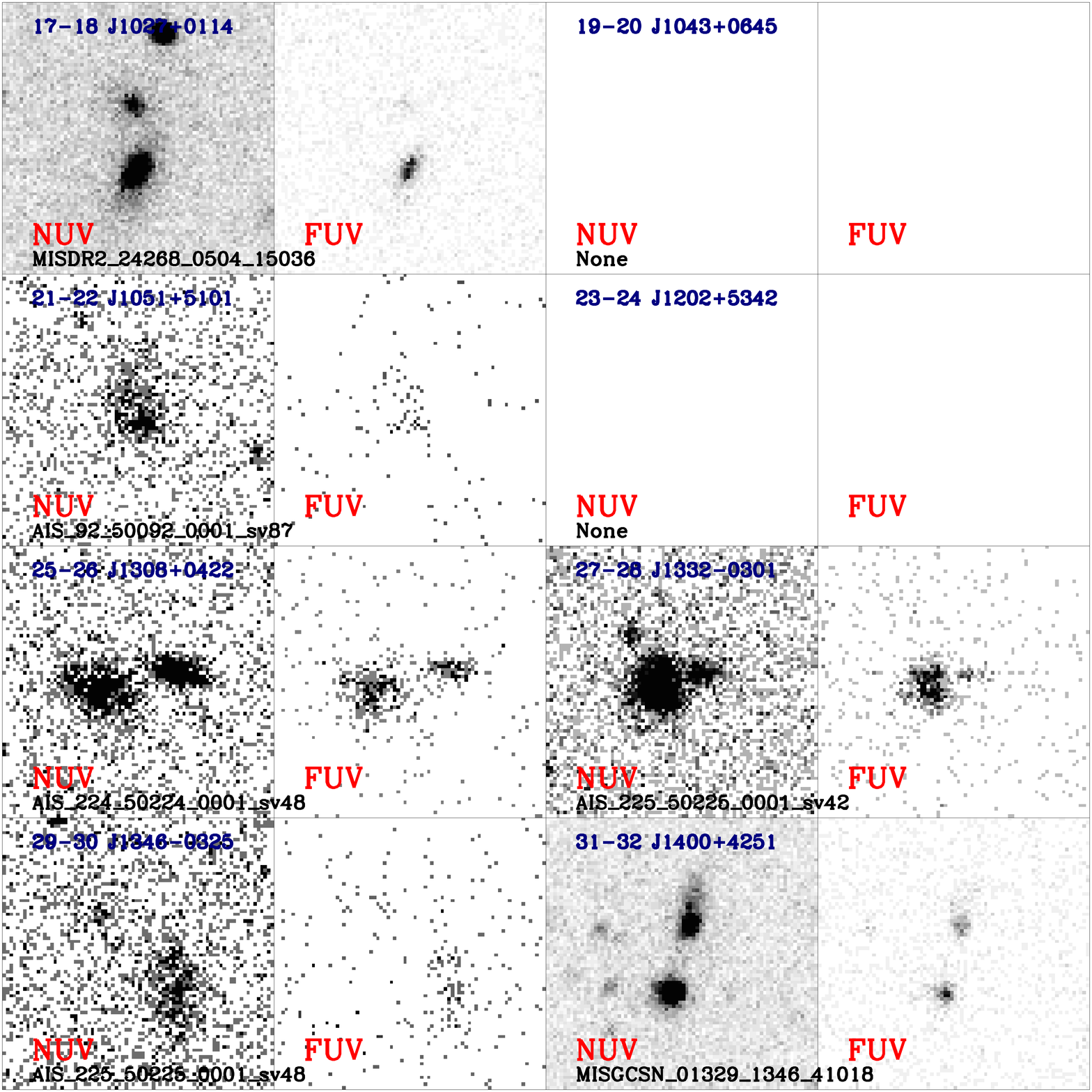}% 
  \caption{(Continued.)}
\end{figure*}

\addtocounter{figure}{-1} 
\begin{figure*}
  \centering
  \includegraphics[width=0.7\textwidth]{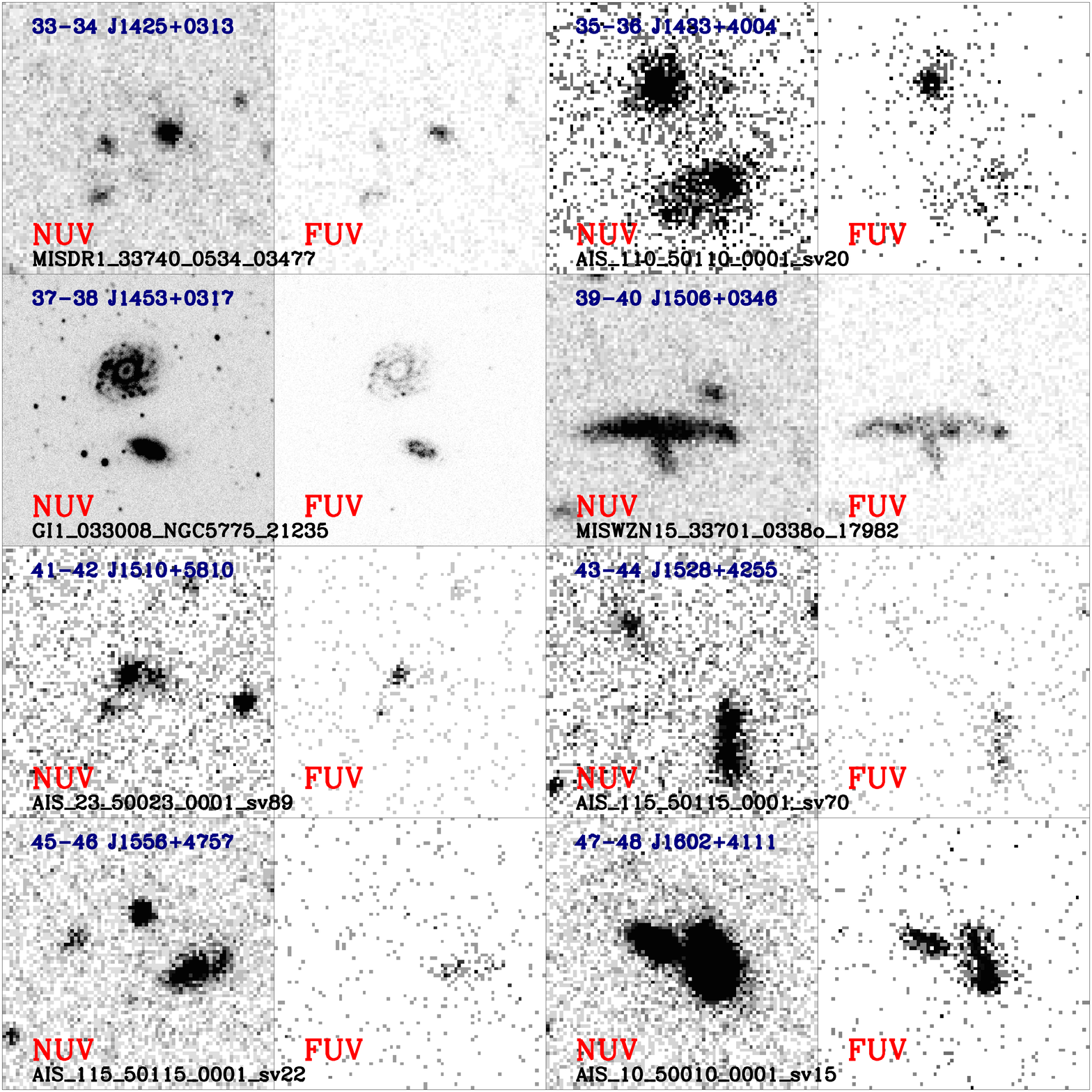}% 
  \caption{(Continued.)}
\end{figure*}

\addtocounter{figure}{-1} 
\begin{figure*}
  \centering
  \includegraphics[width=0.7\textwidth]{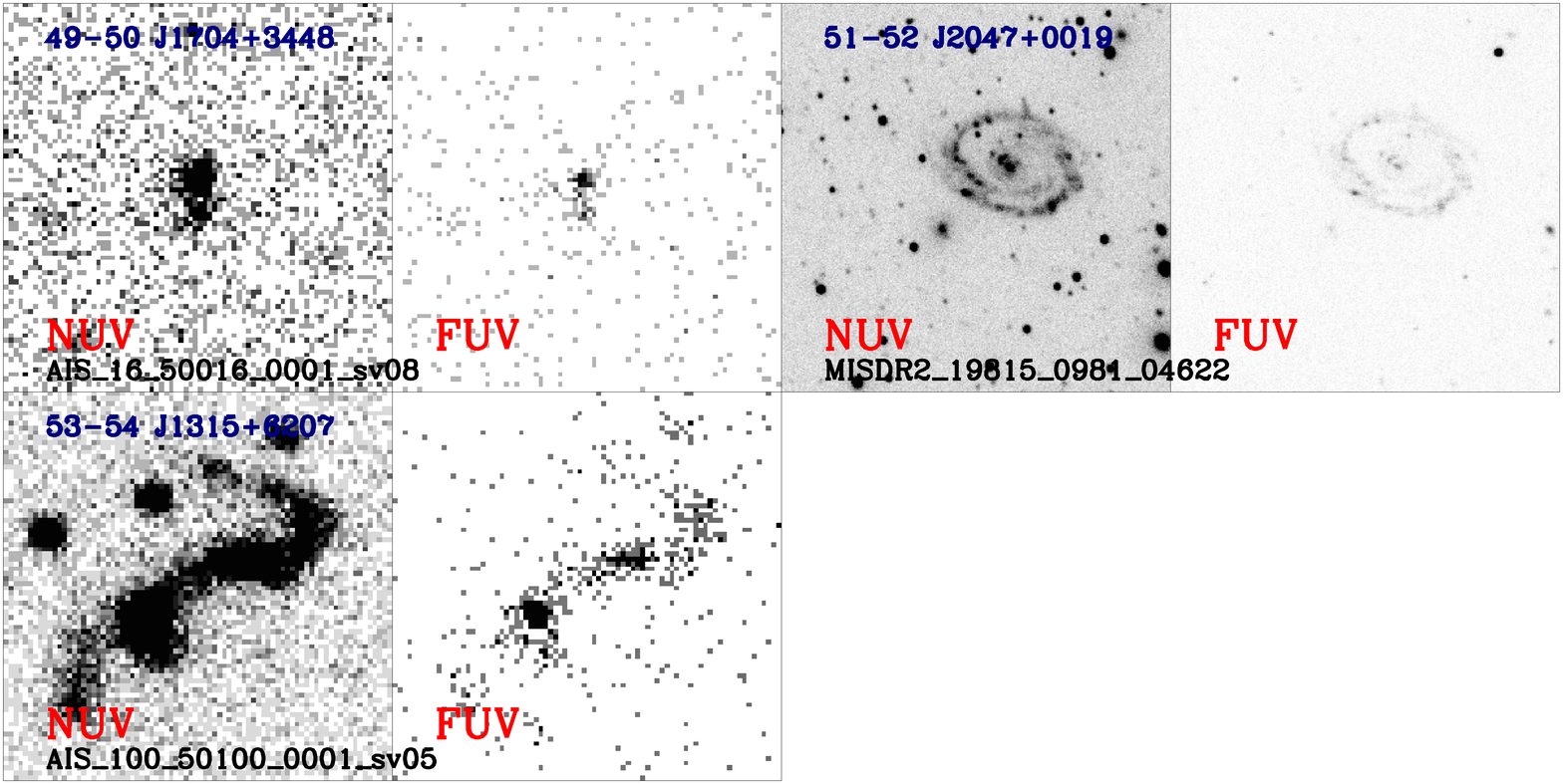}% 
  \caption{(Continued.)}
\end{figure*}

\end{appendix}

\end{document}